\documentclass[12pt]{article}
\usepackage{amsfonts}
\usepackage{amssymb}
\usepackage{graphicx}
\usepackage{amsmath}
\usepackage[bottom]{footmisc}
\usepackage{xcolor}
\usepackage[longnamesfirst]{natbib}
\usepackage{rotating}
\usepackage{bm}

\newcommand{\bc}{\begin{center}}
\newcommand{\ec}{\end{center}}
\newcommand{\be}{\begin{equation}}
\newcommand{\ee}{\end{equation}}
\newcommand{\bea}{\begin{eqnarray}}
\newcommand{\eea}{\end{eqnarray}}
\newcommand{\bean}{\begin{eqnarray*}}
\newcommand{\eean}{\end{eqnarray*}}

\newtheorem{algo}{Algorithm}

\newcounter{pkt}

\def\EE{\mathord{I\kern-.35em E}}
\def\PP{\mathord{I\kern-.3em P}}
\def\QQ{\mathord{Q\kern-5pt\hbox{\raise1.1pt\hbox{\vrule height5pt}}\kern5pt}}
\def\RR{\mathord{I\kern-.3em R}}

\newcounter{saveeqn}

\begin{document}

\title{\vspace*{-0.5 in} \textbf{Sequential Monte Carlo With Model Tempering}}

\author{Marko Mlikota\thanks{
		\setlength{\baselineskip}{4mm}  Correspondence:  Department of Economics, University of Pennsylvania, 133 South 36th Street, Philadelphia, PA 19104-6297. Email: mlikota@sas.upenn.edu (Mlikota) and schorf@ssc.upenn.edu (Schorfheide). Schorfheide gratefully acknowledges financial support from the National Science Foundation under Grant SES 1851634.} \\
	{\em \small University of Pennsylvania}
	\and Frank Schorfheide \\
	{\em \small University of Pennsylvania}, \\ {\em \small CEPR, PIER, NBER} }

\date{This Version: \today}
\maketitle

\begin{abstract}
Modern macroeconometrics often relies on time series models for which it is time-consuming to evaluate the likelihood function. We demonstrate how Bayesian computations for such models can be drastically accelerated by reweighting and mutating posterior draws from an approximating model that allows for fast likelihood evaluations, into posterior draws from the model of interest, using a sequential Monte Carlo (SMC) algorithm.  We apply the technique to the estimation of a vector autoregression with stochastic volatility and a nonlinear dynamic stochastic general equilibrium model. The runtime reductions we obtain range from 27\% to 88\%. (JEL C11, C32)
\end{abstract}

\noindent {\footnotesize  {\em Key words:} Bayesian Computations, Dynamic Stochastic General Equilibrium Models, Sequential Monte Carlo, Stochastic Volatility, Vector Autoregressions.}

\thispagestyle{empty}

\clearpage
\setcounter{page}{1}

\section{Introduction}
\label{sec:introduction}

Modern macroeconometrics often relies on time series models for which it is time-consuming to evaluate the likelihood function, either because it takes a long time to solve the underlying structural model, or the likelihood evaluation requires to integrate out latent state variables. In this paper we demonstrate how Bayesian computations for such models can be accelerated by reweighting and mutating posterior draws from an approximating model that allows for fast likelihood evaluations. We show that a sequential Monte Carlo (SMC) algorithm that starts out with draws from the posterior distribution of an approximating model instead of the prior distribution of the target model can drastically speed up the posterior computations.

SMC methods have been traditionally used to solve nonlinear filtering problems, an example being the bootstrap particle filter of \cite{GordonSalmondSmith1993}. Subsequently, \cite{Chopin2002} showed how to adapt particle filtering techniques to conduct posterior inference for a static parameter vector. The first paper that applied SMC techniques to posterior inference for the parameters of a (small-scale) DSGE model was \cite{Creal2007}. Subsequent work by \cite{herbst2014sequential,HerbstSchorfheide2016} fine-tuned the algorithm so that it could be used for the estimation of medium- and large-scale models.
\cite{DurhamGeweke2011} show how to parallelize a flexible and self-tuning SMC algorithm for the estimation of time series models on graphical processing units (GPU).

In general, SMC algorithms approximate a target posterior distribution by creating intermediate approximations to a sequence of bridge distributions, indexed in this paper by $n$. At each stage, the current bridge distribution is represented by a swarm of so-called particles. Each particle is composed of a value and a weight. Weighted averages of the particle values converge to expectations under the stage-$n$ distribution. The transition from stage $n-1$ to $n$ involves changing the particle weights and values (mutation) so that the swarm adapts to the new distribution. Typically, these bridge distributions are constructed by either using the full-sample likelihood ({\em likelihood tempering}, LT)---generated by raising this likelihood function to the power of $\phi_n$, where $\phi_n$ increases from zero to one---or by sequentially adding observations to the likelihood function ({\em data tempering}, DT).

As initially suggested in \cite{CaiEtAl2020} but not explored any further, in this paper we document the runtime reductions achievable by a {\em model tempering approach} that takes a geometric average with weights $\phi_n$ and $1-\phi_n$ of the likelihood functions associated with the target model, denoted by $M_1$, and an approximating model $M_0$.\footnote{\cite{AcharyaEtAl2021} discuss model tempering as a strategy to estimate a HANK model.} Building on earlier work in the statistics literature, e.g., \cite{JasraStephensDoucetTsagaris2011}, and work in the DSGE model literature, e.g., \cite{herbst2019tempered} and Cai {\em et al.} (2021), we choose the tempering schedule defined through the $\phi_n$ sequence adaptively. Our adaptive schedules are calibrated by a single tuning parameter that controls the desired variance of the particle weights. The smaller the discrepancy between the posterior distribution of the approximating and the original model, the fewer bridge distributions are being used, and the faster the posterior analysis.

In general, model tempering is an attractive computational strategy for applications in which the likelihood evaluation for the target model is computationally costly and there is an approximating model for which the likelihood evaluation is fast and generates a posterior that is not too different from the posterior of the target model. We envision the approximating model to be a simplified version of the target model for which posterior computations are also implemented via SMC, in this case with likelihood tempering.\footnote{Even in the absence of a model tempering strategy, estimating approximating models is desirable as part of the modeling and code debugging that ultimately leads to the target model.} The $M_0$ likelihood tempering can be terminated before the weight on the likelihood function has reached the value one. We denote the terminal weight on the $M_0$ likelihood by $\psi_* \in (0,1]$. The early termination will lead to a more diffuse $M_0$ posterior, draws from which might be more easily mutable into draws from the $M_1$ posterior in the subsequent model tempering steps. This feature introduces additional flexibility into the model tempering algorithm.

We provide a formula for the runtime reduction achievable by model tempering that depends on the number of stages as a function of $\psi_*$ used for the $M_0$ and $M_1$ SMC runs, respectively, and the relative time it takes to evaluate the likelihood functions of the two models, denoted by the ratio $\tau_0/\tau_1$. Note that the user can evaluate $\tau_0/\tau_1$ before running the entire algorithm. We show that the runtime reduction profile is convergent as $\tau_0/\tau_1 \longrightarrow 0$. In the limit, the runtime reduction is determined just by the number of $M_0$ and $M_1$ SMC stages, which in turn depends on the alignment of the $\psi_*$-tempered $M_0$ posterior and the target $M_1$ posterior, relative to the alignment of the prior and the $M_1$ posterior. To assess the potential gains of model tempering {\em ex ante}, we recommend that the researcher computes the variance of the importance sampling weights, that would be needed to reweight the draws from the $\psi_*$-tempered $M_0$ posterior to approximate the target $M_1$ posterior, for various choices of $\psi_*$. If there is a $\psi_*$ for which this variance is small relative to the number of SMC particles, then the gains from model tempering are potentially large.

We consider three numerical illustrations of model tempering. In the first illustration, both target and approximating densities are univariate Normal. We illustrate how the distance between the densities affects the number of stages (and computational time) required to convert draws from the approximating density into draws from the target density. In the second illustration we consider the estimation of a vector autoregression (VAR) with stochastic volatility (SV), using a homoskedastic VAR as approximating model. In our illustration, model tempering is able to reduce the computational time by 79\%. At last we consider the estimation of a dynamic stochastic general equilibrium (DSGE) model. We take $M_1$ as a version of the model that is solved with a second-order perturbation around the steady state and for which the likelihood function is evaluated with a bootstrap particle filter (BSPF). The approximating model is a log-linearized version for which the likelihood function can be evaluated quickly using the Kalman filter. In our numerical example, model tempering can reduce the runtime of our JULIA code from 655 to 80 minutes.

The remainder of this paper is organized as follows. Section~\ref{sec:smc} describes the proposed model tempering SMC algorithm. Section~\ref{sec:example} considers the simple example based on univariate Gaussian posterior distributions. In Section~\ref{sec:VARSV} we use model tempering to estimate the VAR with SV. In Section~\ref{sec:DSGE} we implement our algorithm for the nonlinear DSGE model. Section~\ref{sec:conclusion} concludes. An Online Appendix contains supplemental information on the methodology and further details and results for the numerical illustrations.

\section{Bayesian Inference, SMC, and Model Tempering}
\label{sec:smc}

VARs, DSGE models, and other time series models are often estimated using Bayesian inference for several reasons. First, the Bayesian framework provides a powerful toolkit to handle the presence of latent variables in state-space models. Second, uncertainty about parameters, shocks, and unobserved state variables is treated identically which makes it conceptually straightforward to form predictive distributions that reflect all sources of uncertainty. Third, prior distributions can be used to regularize the estimation of high-dimensional models (e.g., VARs) or to incorporate additional information not contained in the estimation sample (DSGE model estimation).

Bayesian inference combines a prior distribution $p(\theta)$ with a likelihood function $p(Y|\theta)$ to form a posterior distribution $p(\theta|Y)$, which is given by
\be
\pi(\theta) \equiv p(\theta|Y) = \frac{p(Y|\theta)p(\theta)}{p(Y)}, \quad p(Y) = \int p(Y|\theta)p(\theta) d\theta,
\ee
where $Y=Y_{1:T}=\{y_1, y_2, ..., y_T\}$ and the normalization constant $p(Y)$ is called the marginal data density (MDD). In most applications, the posterior distribution $p(\theta|Y)$ does not belong to a family of distributions for which moments and percentiles can be easily calculated or draws can be obtained by direct sampling. In this paper we use an SMC algorithm to sample from the posterior distribution $p(\theta|Y)$. The algorithm combines insights from importance sampling and Markov chain Monte Carlo (MCMC) techniques. Two of its key advantages are that it is able to provide accurate approximations of non-regular posterior distributions and that it can be easily parallelized, unlike MCMC algorithms. In Section~\ref{subsec:smc.generic} we describe a generic SMC algorithm to sample from the posterior distribution of $\theta$. The section draws heavily from the more detailed exposition in \cite{herbst2014sequential,HerbstSchorfheide2016}. Model tempering, which is the focus of our paper, is introduced in Section~\ref{subsec:smc.tempering} and implementation details are discussed in Section~\ref{subsec:smc.implementation}. In Section~\ref{subsec:smc.gains} we assess potential runtime reductions.

\subsection{A Generic SMC Algorithm}
\label{subsec:smc.generic}

\begin{figure}[t!]
	\caption{Evolution of Bridge Distributions}
	\label{fig:smc.bridge.distributions}
	\begin{center}
        \includegraphics[width=5in]{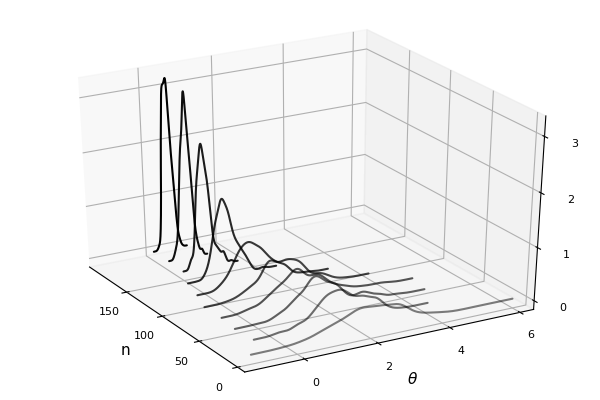}
    \end{center}
     {\footnotesize {\em Notes:} The sequence of bridge distributions for a scalar parameter $\theta$ is shown along the y-axis.}\setlength{\baselineskip}{4mm}
\end{figure}

In order to draw from $\pi(\theta)$, the SMC algorithm uses a sequence of bridge posterior distributions $\{\pi_n(\theta)\}_{n=0}^{N_\phi}$, illustrated in Figure~\ref{fig:smc.bridge.distributions}, where the last one in the sequence equals the posterior distribution -- $\pi_{N_\phi}(\theta)=\pi(\theta)$ -- and where each $\pi_{n-1}(\theta)$ is used as the proposal density for $\pi_n(\theta)$. The bridge posteriors are constructed from stage $n$ likelihood functions $p_n(Y|\theta)$ and defined as
\be
    \pi_n(\theta) = \frac{p_n(Y|\theta) p(\theta)}{\int p_n(Y|\theta) p(\theta) d\theta}.
\ee
Each density $\pi_n(\theta)$ is represented by a particle approximation $\{\theta^i_n,W^i_n\}_{i=1}^N$. Thus, at stage $n$ the algorithm propagates the particles $\{\theta^i_{n-1},W^i_{n-1}\}_{i=1}^N$ so that they come to represent the target density $\pi_n(\theta)$. Formally, the algorithm proceeds in the following steps:

\begin{algo}[Generic SMC Algorithm] \label{algo:smc} \hspace*{1cm}\\[-3ex]
	\begin{enumerate}
		\item {\bf Initialization.} ($n = 0$ and $\phi_{0} = 0$.)
		Draw the initial particles from $\pi_0(\theta)$: $\theta^{i}_{1} \sim \pi_0(\theta)$ and
		$W^{i}_{1} = 1$, $i = 1, \ldots, N$.
		\item {\bf Recursion.} For $n = 1, \ldots, N_{\phi}$,
		\begin{enumerate}
			\item {\bf Correction.}  Reweight the particles from stage $n-1$ by defining
			the incremental weights
			\be
			\tilde w_{n}^{i} = \frac{p_n(Y |\theta^{i}_{n-1})}{p_{n-1}(Y|\theta^{i}_{n-1})}
			\label{eq:smcdeftildew}
			\ee
			and the normalized weights
			\be
			\tilde{W}^{i}_{n} = \frac{\tilde w_n^{i} W^{i}_{n-1}}{\frac{1}{N} \sum_{i=1}^N \tilde w_n^{i} W^{i}_{n-1}}, \quad
			i = 1,\ldots,N.
			\label{eq:smcdeftildebigw}
			\ee
			\item {\bf Selection (Optional).}
			Resample the swarm of particles, $\{\theta_{n-1}^i,\tilde{W}_n^i \}_{i=1}^N$, and denote resampled particles by
			$\{\hat{\theta}_n^i, W_n^i \}_{i=1}^N$, where $W_n^i=1$ for all $i$.

			\item {\bf Mutation.}  Starting from $\hat{\theta}_n^i$, propagate the particles $\{\hat{\theta}_n^i,W_n^i \}$ via $N_{MH}$
			steps of a Metropolis-Hastings (MH) algorithm with transition density $K_n(\theta| \tilde{\theta}; \zeta_n)$ and
			stationary distribution $\pi_n(\theta)$.  Note that the weights are unchanged, and denote the mutated particles by $\{ \theta_n^i, W_n^i \}_{i=1}^N.$
		\end{enumerate}
		An approximation of $\mathbb{E}_{\pi_n}[h(\theta)]$ is given by
		\be
		\bar{h}_{n,N} = \frac{1}{N} \sum_{i=1}^N h(\theta_{n}^i) W^i_n.
		\label{eq:defbarh}
		\ee

		\item For $n=N_{\phi}$ ($\phi_{N_\phi}=1)$ the final importance sampling
		approximation of $\mathbb{E}_\pi[h(\theta)]$ is given by:
		\be
		\bar{h}_{N_\phi,N}  = \sum_{i=1}^{N} h(\theta_{N_\phi}^i) W_{N_\phi}^i.
		\ee
	\end{enumerate}
\end{algo}

The correction step is a classic importance sampling step, in which the particle weights are updated to reflect the stage $n$ distribution $\pi_n(\theta)$. The selection step is optional. On the one hand, resampling adds noise to the Monte Carlo approximation, which is undesirable. On the other hand, it equalizes the particle weights, which increases the accuracy of subsequent importance sampling approximations.  The decision of whether or not to resample is typically based on a threshold rule for the variance of the particle weights which can be transformed into an effective particle sample size (ESS):
\be
\widehat{ESS}_n = N \big/ \left( \frac{1}{N} \sum_{i=1}^N (\tilde{W}_n^i)^2\right).
\label{eq:ESS}
\ee
If the particles have equal weights, then $\widehat{ESS}_n = N$. If one particle has weight $N$ and all other particles have weight 0, then $\widehat{ESS}_n = 1$. These are the upper and lower bounds for the effective sample size. To balance the trade-off between adding noise and equalizing particle weights, we execute the resampling step if $\widehat{ESS}_n$ falls below $N/2$ using a systematic resampling algorithm.

The mutation step changes the particle values. In the absence of the mutation step, the particle values would be restricted to the set of values drawn in the initial stage from the prior distribution. This would clearly be inefficient, because the prior distribution is typically a poor proposal distribution for the posterior in an importance sampling algorithm. As the algorithm cycles through the $N_\phi$ stages, the particle values successively adapt to the shape of the posterior distribution. This is the key difference between SMC and classic importance sampling. The transition kernel $K_n(\theta|\tilde{\theta}; \zeta_n)$ is designed to have the following invariance property:
\be
\pi_n(\theta_n) = \int K_n(\theta_n|\hat{\theta}_n;\zeta_n) \pi_n(\hat{\theta}_n) d\hat{\theta}_n.
\ee
Thus, if $\hat{\theta}_n^i$ is a draw from $\pi_n$, then so is $\theta_n^i$. The mutation step can be implemented by using one or more steps of a MH algorithm. The probability of mutating the particles can be increased by blocking the elements of the parameter vector $\theta$ or by iterating the MH algorithm over multiple steps. The vector $\zeta_n$ summarizes the tuning parameters of the MH algorithm.

\subsection{Model Tempering}
\label{subsec:smc.tempering}

Up to now we imposed minimal conditions on the sequence of bridge posterior distributions. To initialize the algorithm, we implicitly required required that it is possible to sample from the initial distribution $\pi_0(\theta)$, which is typically the prior $p(\theta)$ under likelihood or data tempering, and we required that the stage $N_\phi$ posterior is equal to the target posterior distribution: $\pi_{N_\phi}(\theta) = \pi(\theta)$. While previous applications of Algorithm~\ref{algo:smc} in econometrics focused on either data or likelihood tempering, the contribution of our paper is to assess the performance of the model tempering approach. Under model tempering the bridge distributions are constructed as follows. Let $M_1$ be the target model with likelihood function $p(Y|\theta,M_1)$ and let $M_0$ be an approximating model with likelihood function $p(Y|\theta,M_0)$. We define the bridge likelihood functions that are used in Steps 2(a) and 2(c) of Algorithm~\ref{algo:smc} as:
\be
p_n(Y|\theta) =p(Y|\theta,M_1)^{\phi_n}p(Y|\theta,M_0)^{1-\phi_n} \; , \quad \phi_0=0, \quad \phi_{N_\phi} = 1, \quad \phi_n \uparrow 1.
\label{eq:n.lik.MT}
\ee
It can be easily seen that $\pi_{N_\phi}(\theta) = p(\theta|Y,M_1)$, as required, and that the algorithm is initialized with draws from the $M_0$ posterior $\pi_0(\theta) = p(\theta|Y,M_0)$. The intermediate distributions are obtained by shifting the weight gradually from the $M_0$ posterior to the posterior of the target model $M_1$.

Model tempering distinguishes itself from the two most widely-used tempering schemes, likelihood tempering and data tempering, neither of which involve an approximating model $M_0$. Under likelihood tempering (e.g., \cite{herbst2014sequential}) the stage $n$ posterior is constructed from a tempered version of the full-sample likelihood function:
\[
  p_n(Y|\theta) = p(Y|\theta,M_1)^{\phi_n}.
\]
Under data tempering (e.g., \cite{DurhamGeweke2011}) the bridge distributions are obtained from a fraction of the sample observations
$\pi_n(\theta) \propto p(Y_{1:\lfloor \phi_n T \rfloor}|\theta,M_1)p(\theta)$ or, as in Cai {\em et al.} (2021), by gradually shifting the weight from a short-sample likelihood to a full-sample likelihood:
\[
p_n(Y|\theta) = p(Y_{1:T}|\theta,M_1)^{\phi_n}p(Y_{1:T_0}|\theta,M_1)^{1-\phi_n}, \quad T_0 < T.
\]

Model tempering is a computationally efficient alternative under two conditions. First, the likelihood evaluation of the target model $M_1$ is computationally costly, whereas the likelihood evaluation of the approximating model $M_0$ is, in relative terms, fast. Second, the likelihood functions of the target and the approximating model have to be sufficiently close such that only a modest number of intermediate stages are required to convert draws from the $M_0$ posterior into draws from the $M_1$ posterior. We provide a more detailed discussion in Section~\ref{subsec:smc.gains} below.

\subsection{Implementation Details}
\label{subsec:smc.implementation}

\noindent {\bf Adaptive Tempering Schedule.} Under the adaptive tempering schedule used in Cai {\em et al.} (2021) $\phi_n$ is chosen to target a desired level of the ESS defined in (\ref{eq:ESS}). Emphasizing the dependence of the incremental weights on the current tempering coefficient $\phi$, write $\tilde{w}_n^i$ in (\ref{eq:smcdeftildew}) as
\[
\tilde{w}^i(\phi) = \frac{p(Y|\theta^i_{n-1},M_1)^\phi p(Y|\theta^i_{n-1},M_0)^{1-\phi}}{p(Y|\theta^i_{n-1},M_1)^{\phi_{n-1}} p(Y|\theta^i_{n-1},M_0)^{1-\phi_{n-1}}}
\]
and define
\[
f(\phi) = \widehat{ESS}_n(\phi) - \alpha \widehat{ESS}^*_{n-1}, \quad 0 < \underline{\alpha} \le \alpha < 1,
\]
where $\widehat{ESS}^*_{n-1} = \widehat{ESS}_{n-1}$ if the stage $n-1$ selection step (resampling) was executed and $\widehat{ESS}^*_{n-1} = N$ otherwise. Let $\phi_n^*$ satisfy $f(\phi_n^*)=0$ and define $\phi_n = \min \{\phi_n^*,1\}$.

The parameter $\alpha$, to be specified by the user, is the targeted reduction in ESS. It can be shown that for $\phi > \phi_{n-1}$ the ESS satisfies the inequality $\widehat{ESS}_n(\phi) < \widehat{ESS}^*_{n-1}$. Moreover, $\widehat{ESS}_n(\phi)$ is a strictly decreasing function of $\phi$ such that $f(\phi)=0$ has a unique solution. The adaptive algorithm chooses the tempering schedule to control the deterioration of the ESS statistic. The smaller the slope of the function $\widehat{ESS}_n(\phi)$, the larger the increments in the tempering schedule.  The number of stages $N_\phi$ is then endogenously determined and is equal to the stage $n$ at which $\phi_n = 1$.

\noindent {\bf Model-Specific Parameters.} It might be the case that not all of the parameters that appear in $M_1$ also affect $M_0$, or vice versa. For instance, in one of our illustrations, $M_1$ is a VAR with SV, whereas $M_0$ is a homoskedastic VAR. Thus, the $M_1$ parameter vector contains additional parameters that govern the dynamics of the SV processes. Partition $\theta' = [\theta_c',\theta_0',\theta_1']$, where $\theta_c$ is the vector of common parameters and $\theta_j$ are parameters specific to model $M_j$. The likelihood functions are given by
\[
   p(Y|\theta,M_j) = p(Y|\theta_c,\theta_j,M_j), \quad j=0,1
\]
and
\be
   p_n(Y|\theta) = p(Y|\theta_c,\theta_1,M_1)^{\phi_n}
                   p(Y|\theta_c,\theta_0,M_0)^{1-\phi_n}.
\ee
Consider stage $n=0$ with $\phi_0=0$. Because $\theta_1$ does not enter the $M_0$ likelihood function, its distribution does not get updated in view of the data $Y$ and we can factorize the $M_0$ posterior as follows.
\[
 \pi_0(\theta) = p(\theta|Y,M_0)
  = p(\theta_c,\theta_0|Y,M_0) p(\theta_1).
\]
Thus, the model tempering SMC algorithm starts from posterior draws of $(\theta_c,\theta_0)$ and prior draws from $\theta_1$. The use of prior draws for $\theta_1$ in the absence from any information through $M_0$ is both natural and desirable.

At stage $n=N_\phi$ the SMC algorithm approximates the $M_1$ posterior which, on the enlarged parameter space, is given by
\[
\pi_{N_\phi}(\theta) = p(\theta|Y,M_1)
= p(\theta_c,\theta_1|Y,M_1) p(\theta_0).
\]
The ultimate object of interest is, in slight abuse of notation, the marginal posterior
\[
   p(\theta_c,\theta_1|Y,M_1)=\int \pi_{N_\phi}(\theta_c,\theta_0,\theta_1) d\theta_0
   .
\]
While the SMC sampler generates draws from the joint posterior of $(\theta_c,\theta_0,\theta_1)$, draws from the marginal posterior can be obtained by simply dropping the $\theta_0$ draws. A potential disadvantage of including $\theta_0$ into the definition of $\theta$ is that the SMC algorithm has to turn $\theta_0$ draws from a potentially highly concentrated posterior $p(\theta_0|Y,M_0)$ into draws from a more diffuse prior $p(\theta_0)$, which may require an undesirably large number of steps. Thus, we recommend to simply fix $\theta_0$ at a reasonable value, e.g., the posterior mean or mode from a preliminary estimation of $M_0$, and then drop it from the definition of $\theta$.

\noindent {\bf Marginal Data Density Ratio.} The SMC algorithm produces as a by-product an approximation of the marginal likelihood ratio $p(Y|M_1)/p(Y|M_0)$. Note that
\be
\frac{1}{N} \sum\limits_{i=1}^{N} \tilde{w}^i_n \tilde{W}^i_{n-1}  \approx \int \frac{p_n(Y|\theta)}{p_{n-1}(Y|\theta)} \left[\frac{p_{n-1}(Y|\theta) p(\theta)}{\int p_{n-1}(Y|\theta) p(\theta) d\theta} \right] d\theta
= \frac{\int p_n(Y|\theta) p(\theta) d\theta}{\int p_{n-1}(Y|\theta) p(\theta) d\theta}, \label{eq:mdd.heuristic}
\ee
where
\begin{eqnarray*}
	\int p_{0}(Y|\theta) p(\theta)d\theta &=& \int p(Y|\theta,M_0)p(\theta)d\theta = p(Y|M_0) \\
	\int p_{N_\phi}(Y|\theta) p(\theta)d\theta &=& \int p(Y|\theta,M_1)p(\theta)d\theta = p(Y|M_1).
\end{eqnarray*}
In turn, it can be shown that
\be
\prod_{n=1}^{N_\phi} \left( \frac{1}{N} \sum_{i=1}^N \tilde{w}_n^i W_{n-1}^i \right) \;
\stackrel{\mbox{a.s.}}{\longrightarrow} \; \frac{p(Y|M_1)}{p(Y|M_0)}
\label{eq:mdd.phat}
\ee
as the number of particles $N \longrightarrow \infty$;  see, for instance, \cite{herbst2014sequential}.

\noindent {\bf Tempered $M_0$ Posterior.} Rather than using the full-information posterior under $M_0$ as the proposal density, one can choose to incorporate only a fraction of the information embodied in the posterior under model $M_0$. Suppose that the draws from $M_0$ are generated through an SMC algorithm with likelihood tempering, which is what we are doing in the illustrations in Sections~\ref{sec:VARSV} and~\ref{sec:DSGE}. Then we can define
\be
  p_n(Y|\theta) =p(Y|\theta,M_1)^{\phi_n}\big[p(Y|\theta,M_0)^{\psi_*}\big]^{1-\phi_n}, \quad \psi_* \in [0,1) \;,
\ee
which leads to the initialization
\be
\pi_0(\theta;\psi_*) \propto p(Y|\theta,M_0)^{\psi_*}p(\theta) \; .\label{eq:pi0_M1}
\ee
The density $\pi_0(\theta;\psi_*)$ represents the posterior obtained from the tempered $M_0$ likelihood function. Thus, the posterior sampling for the approximating model is terminated at $\phi_{N_\phi}=\psi_* < 1$ instead of $\phi_{N_\phi}=1$. The advantage of this strategy is that for $\psi_*<1$ the density $\pi_0(\theta;\psi_*)$ is more diffuse than the full $M_0$ posterior and may exhibit a greater overlap with the target posterior in applications in which $M_0$ and $M_1$ posteriors differ substantially. Note that for $\psi_*=0$ the model $M_1$ would be estimated by standard likelihood tempering instead of model tempering.

\subsection{Computational Gains}
\label{subsec:smc.gains}

To formalize the discussion of the computational advantage of model tempering we begin by introducing some additional notation. Let $\tilde{N}_0(\psi_*) = N_\phi^0(\psi_*)+1$ be the number of $M_0$ SMC stages to obtain a particle swarm that approximates $\pi_0(\theta;\psi_*)$ in (\ref{eq:pi0_M1}), including the initial stage, which draws from the prior $p(\theta)$. For the subsequent $M_1$ model tempering we write the number of stages as $\tilde{N}_1(\psi_*) = N_\phi^1(\psi_*)+1$, again to emphasize the dependence on $\psi_*$. We regard $\psi_*=0$ as $M_1$ likelihood tempering and adopt the convention that $\tilde{N}_0(0)=0$.

In the typical VAR and DSGE model applications for which the model tempering procedure is developed, the runtime of the SMC algorithm is predominantly determined by the time it takes to evaluate the likelihood function of the underlying models. Let $N_*$ be the number of likelihood evaluations per SMC stage. It is given by
$N_* = N \cdot N_{MH} \cdot N_{blocks}$, where $N$ is the number of particles, $N_{MH}$ is the number of Metropolis-Hastings (MH) steps during the mutation phase, and $N_{blocks}$ is the number of parameter blocks used in each MH step. Moreover, let $\tau_j$ be the time it takes to evaluate the likelihood function of model $M_j$, $j=0,1$. Then the total runtime is given by
\be
	  {\cal T}(\psi_*,\tau_1,\tau_0)
= 	N_* \big( \tilde{N}_1(\psi_*)\tau_1 + \mathbb{I}\{\psi_*>0\} (\tilde{N}_1(\psi_*)+ \tilde{N}_0(\psi_*)) \tau_0  \big)
 ,
\label{eq:smc.runtime.calT}
\ee
where $\mathbb{I}\{x > a\}$ is the indicator function that is equal to one if $x>a$ and equal to zero otherwise.
Under likelihood tempering, i.e., $\psi_*=0$, the likelihood function of $M_1$ has to be evaluated $N_* \tilde{N}_1(0)$ times and there is no need to evaluate the $M_0$ likelihood. Under model tempering with $\psi_* > 0$, the likelihood function of $M_1$ has to be evaluated $\tilde{N}_1(\psi_*)$ times and the likelihood of $M_0$ needs to be evaluated during the $M_0$ SMC run and the $M_1$ SMC run.

As mentioned in Section~\ref{subsec:smc.tempering}, we are concerned with the case in which the use of the (tempered) $M_0$ posterior reduces the number stages for the $M_1$ SMC and the $M_1$ likelihood evaluation is substantially faster than the $M_0$ evaluation:
\[
   \tilde{N}_1(\psi_*) < \tilde{N}_1(0) \; \mbox{for} \; \psi_* > 0 \quad \mbox{and} \quad \tau_0 < \tau_1.
\]
In the numerical illustrations in Sections~\ref{sec:VARSV} and~\ref{sec:DSGE} we report runtimes of model tempering relative to likelihood tempering:\footnote{We found that this formula approximates the actual runtime reductions well.}
\be
{\cal R} (\psi_*,\tau_0/\tau_1)
= \frac{{\cal T}(\psi_*,\tau_1,\tau_0)}{{\cal T}(0,\tau_1,\tau_0)}
= \frac{\tilde{N}_1(\psi_*)}{\tilde{N}_1(0)} + \mathbb{I}\{ \psi_* > 0\} \frac{\tilde{N}_1(\psi_*)+\tilde{N}_0(\psi_*)}{\tilde{N}_1(0)} \frac{\tau_0}{\tau_1}.
\label{eq:smc.runtime.calR}
\ee

The first ratio on the right-hand side of (\ref{eq:smc.runtime.calR}) captures the effect of reducing the number of $M_1$ SMC stages needed to reach the target posterior by starting from the tempered $M_0$ posterior $\pi_0(\theta;\psi_*)$ instead of the prior $p(\theta)$. It does not depend on the relative runtime of the $M_1$ and $M_0$ likelihood evaluations. The second term captures the relative costs of having to evaluate the $M_0$ likelihood function. If $\tilde{N}_1(\psi_*)+\tilde{N}_0(\psi_*) \approx \mbox{const}$ as a function of $\psi_*$, then the second term generates a level shift of ${\cal R} (\psi_*,\tau_0/\tau_1)$.
As the likelihood evaluation of $M_0$ becomes costless relative to the $M_1$ likelihood evaluation,
\be
\lim_{(\tau_0/\tau_1) \longrightarrow 0} \;
{\cal R} (\psi_*,\tau_0/\tau_1) =  \frac{\tilde{N}_1(\psi_*) }{\tilde{N}_1(0)}.
\ee
In the limit, the time it takes to estimate $M_0$ becomes irrelevant and the reduction is purely driven by the reduction in the number of SMC stages resulting from using an initial distribution that is closer to the target posterior.


\begin{figure}[t!]
	\caption{Example: Theoretical Runtime Reductions}
	\label{fig:smc.runtimereductions}
	\begin{center}
    \begin{tabular}{cc}
    	SMC Stages & ${\cal R} (\psi_*,\tau_0/\tau_1)$ \\
		\includegraphics[width=2.8in]{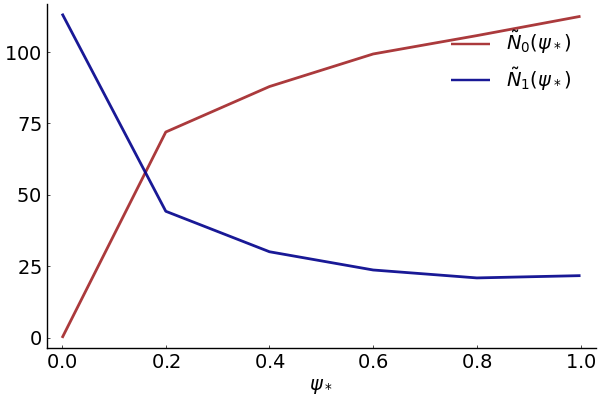} &
		\includegraphics[width=2.8in]{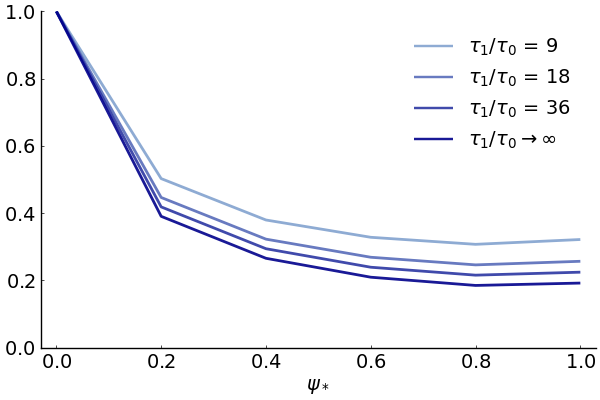}
    \end{tabular}
    \end{center}
	{\footnotesize {\em Notes:} The left panel shows the functions $\tilde{N}_0(\psi_*)$ and $\tilde{N}_1(\psi_*)$, obtained from DGP~1 of Illustration~2 in Section~\ref{sec:VARSV}. The right panel depicts ${\cal R} (\psi_*,\tau_0/\tau_1)$ in (\ref{eq:smc.runtime.calR}).}\setlength{\baselineskip}{4mm}
\end{figure}

In Figure~\ref{fig:smc.runtimereductions} we provide a numerical example for the runtime reduction. In the left panel, we plot functions $\tilde{N}_0(\psi_*)$ and $\tilde{N}_1(\psi_*)$ which are obtained from DGP~1 of the VAR-SV illustration in Section~\ref{sec:VARSV}. The functions are evaluated at $\psi_* \in \{0.0, 0.2, 0.4, 0.6, 0.8, 1.0\}$. As $\psi_*$ increases, the number of stages used in the $M_0$ SMC, denoted by $\tilde{N}_0(\psi_*)$, rises, whereas the number stages in the $M_1$, $\tilde{N}_1(\psi_*)$, falls. The total number of stages required to reach the target posterior stays approximately constant.

The right panel of Figure~\ref{fig:smc.runtimereductions} depicts ${\cal R} (\psi_*,\tau_0/\tau_1)$ for various choices of $\tau_0/\tau_1$. In this example the most runtime drastic reduction occurs by moving from likelihood tempering to $\psi_*=0.2$. For $\psi_* \ge 0.6$ the function is essentially flat. In the VAR illustration $\tau_0/\tau_1=1/9$. We reduce the likelihood-evaluation ratio all the way to 0. The figure indicates that the reduction in the ratio creates a modest downward shift of the level because the sum $\tilde{N}_1(\psi_*)+\tilde{N}_0(\psi_*)$ is fairly insensitive to $\psi_*$.

Thus far, we have provided an {\em ex post} evaluation of computational gains that relied on knowing how the number of SMC stages depends on $\psi$ through the functions $\tilde{N}_1(\psi_*)$ and $\tilde{N}_0(\psi_*)$. To conduct an {\em ex ante} assessment, we recommend the researcher first assesses the times $\tau_j$ it takes to evaluate the likelihood function of the two models. Moreover, we recommend for several values of $\psi_*$ to compute the variance (across $i$) of the importance sampling weights
\be
	\tilde{W}^i(\psi_*) = \frac{\tilde{w}^i(\psi_*)}{\frac{1}{N} \sum_{i=1}^N \tilde{w}^i(\psi_*)}, \quad \tilde{w}^i(\psi_*) = \frac{p(Y|\theta^i,M_1)}{p(Y|\theta^i,M_0)^{\psi_*}},
	\label{eq:defWpsistar}
\ee
where the $\theta^i$'s are draws from $\pi_0(\theta) \propto p(Y|\theta,M_0)^{\psi_*} p(\theta)$. If there is a $\psi_*>0$ for which this variance is considerably smaller than for the $\psi_*=0$ (prior) weights, then there is potential for a substantial runtime reduction. We further explore the relationship between importance sample and runtime reductions in the context of the VAR and DSGE illustrations in Sections~\ref{subsec:VARSV.results} and~\ref{subsec:DSGE.results}.

\section{Illustration 1: Univariate Normal Posteriors}
\label{sec:example}

In the first numerical illustration, we consider an environment in which we can directly control the discrepancy between the approximate posterior and the target posterior. We examine the performance of the model tempering approach as a function of the discrepancy between the posteriors. Starting points are ``posterior'' densities $p(\theta|Y,M_0)$ (approximate) and $p(\theta|Y,M_1)$ (target). We assume that $\theta$ is scalar and approximate and target density are both Normal. In particular, we hold the target density fixed at $p(\theta|Y,M_1) \sim N(0,1)$ and consider a family of approximating densities $p(\theta|Y,M_1) \sim N(\mu,\sigma^2)$, where $\mu$ ranges from -3 to 0 in 0.5 increments and $\sigma$ ranges from 0.2 to 2 in 0.2 increments.

Because in this example we do not construct the posterior density explicitly from a prior distribution and a likelihood function, we let\footnote{This is a slight abuse of notation because $p_n(\theta|Y)$ is not a properly normalized density of $\theta$.}
\[
 p_n(\theta|Y) = p(\theta|Y,M_1)^{\phi_n} p(\theta|Y,M_0)^{1-\phi_n}
\]
and define the incremental weight $\tilde{w}_n^i$ in (\ref{eq:smcdeftildew}) as
\[
\tilde{w}_n^i = \frac{p_n(\theta_{n-1}^i|Y)}{p_{n-1}(\theta_{n-1}^i|Y)}.
\]
We run the SMC Algorithm~\ref{algo:smc} with $N=1,000$ particles, $\alpha = 0.95$, and use $N_{MH}=1$ iteration of a single-block random walk Metropolis-Hastings (RWMH) algorithm in the mutation step with $c_0=0.5$, targeting an acceptance probability of 0.25. The implementation of the mutation step is described in more detail in the Online Appendix.

\begin{figure}[t!]
	\caption{From Approximate to Target Posterior: $p_n(\theta|Y)$}
	\label{fig:example.posteriors}
 	\begin{center}
		\begin{tabular}{ccc}
			$n=0$, $\phi_0=0$ & $n=40$, $\phi_{40} = 0.90$ & $n=71$, $\phi_{71}=1$ \\
            \includegraphics[width=2.0in]{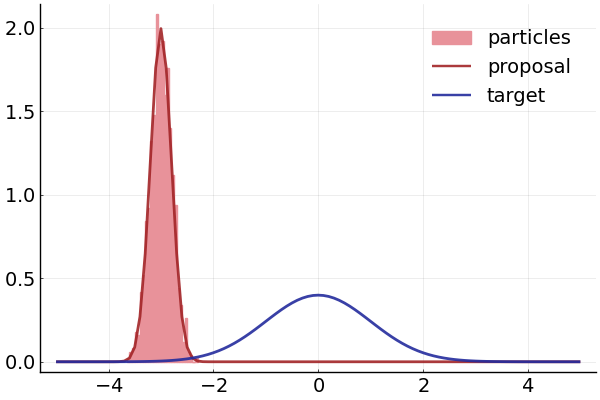} &
            \includegraphics[width=2.0in]{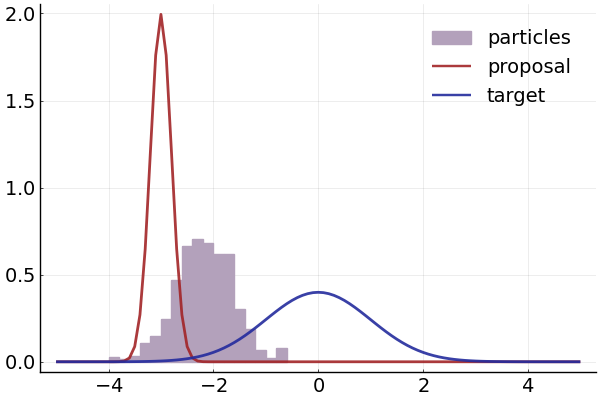} &
            \includegraphics[width=2.0in]{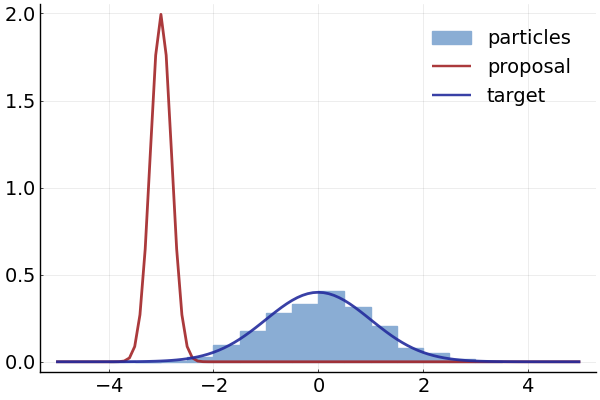}
		\end{tabular}
	\end{center}
	{\footnotesize {\em Notes:} Target density is $N(0,1)$ and approximate density is $N(-3,0.2)$. }\setlength{\baselineskip}{4mm}
\end{figure}

Figure~\ref{fig:example.posteriors} illustrates how the particle swarm moves from an approximate posterior to the target posterior, despite very little overlap between the two densities. We overlay the target posterior density, $N(0,1)$, an approximate density that is used in this example to initialize the algorithm, $N(-3,0.2)$, and a histogram constructed from the stage $n$ particle swarm.  For $n=0$ (left panel) the particle swarm represents the approximate posterior $p(\theta|Y,M_0)$, and for $n=N_\phi=71$ (right panel) it represents the target posterior density $p(\theta|Y,M_1)$. In the center panel of the figure we consider the value of $n=40$ for which the particle swarm represents a weighted geometric mean of the two densities with $\phi_{40}=0.9$.

\begin{figure}[t!]
	\caption{Performance of SMC Algorithm}
	\label{fig:example.performance}
	\begin{center}
		\begin{tabular}{cc}
			{\small (1,1) Average Number of Stages $N_\phi$}
			& {\small (1,2) Tempering Schedule, $\mu=-1$} \\
			\includegraphics[width=2.8in]{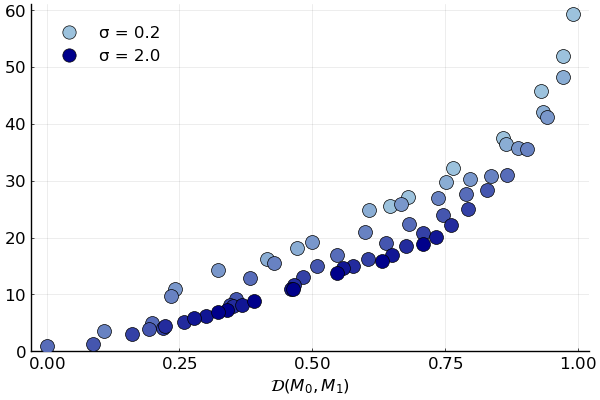} &
			\includegraphics[width=2.8in]{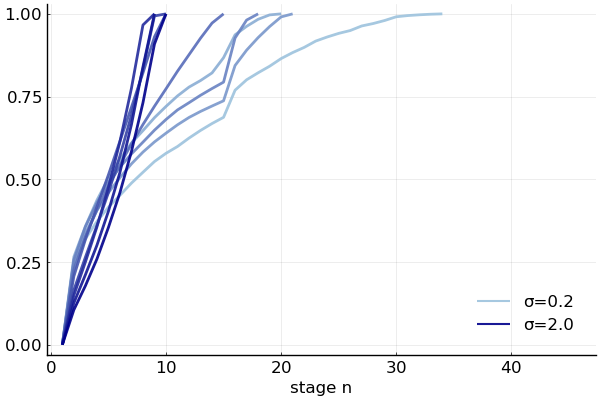} \\[1ex]
			{\small (2,1) Average Runtime [s]} &
			{\small (2,2) Standard Deviation of $\widehat{\mathbb{E}}[\theta|Y,M_1]$} \\
			\includegraphics[width=2.8in]{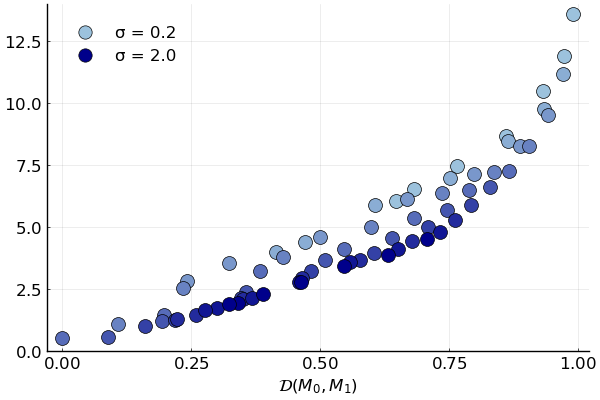} &
			\includegraphics[width=2.8in]{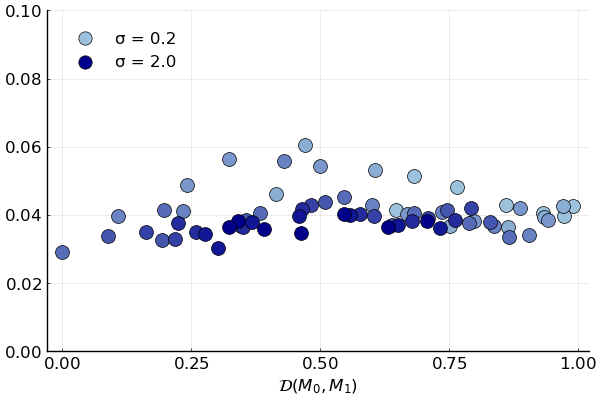}
		\end{tabular}
	\end{center}
	{\footnotesize {\em Notes:} Target density is $N(0,1)$ and approximate densities are $N(\mu,\sigma^2)$ where $\mu$ ranges from -3 to 0 in 0.5 increments and $\sigma$ ranges from 0.2 to 2 in 0.2 increments. The statistics panels (1,1) and (2,1) are averaged across $N_{run}=100$ runs of the SMC algorithm. The standard deviation of the target posterior mean in (2,2) is also computed across multiple runs of the SMC algorithm. In Panel (1,2) we plot the tempering schedules for a single SMC run. Shades of blue indicate different $\sigma$ values.}\setlength{\baselineskip}{4mm}
\end{figure}

In Figure~\ref{fig:example.performance} we illustrate the performance of the SMC algorithm across $N_{run}=100$ runs for the different choices of the approximate posterior. To graphically present the results, we mapped $(\mu,\sigma)$ into a discrepancy measure. In principle one could use the variance of the importance weights defined in (\ref{eq:defWpsistar}). However, it turns out that we consider fairly large discrepancies between $M_0$ and $M_1$ for which the population variance of the importance weights is infinite. Thus, in this section we use an alternative discrepancy measure defined as one minus the area under the minimum of the two densities:
\[
  {\cal D}(M_0,M_1) = 1- \int \min \, \big\{ p(\theta|Y,M_0),\, p(\theta|Y,M_1) \big\} d\theta.
\]
By construction $0 \le {\cal D}(M_0,M_1) \le 1$.

Panel (1,1) shows the average number of stages, $N_\phi$, as a function of ${\cal D}$. In general, the smaller the discrepancy ${\cal D}(M_0,M_1)$, the lower $N_\phi$. Because multiple combinations of $(\mu,\sigma)$ can lead to the same ${\cal D}(M_0,M_1)$, the graph associates multiple $N_\phi$ values with particular values of the discrepancy. For the same level of overlap, approximate densities with a larger standard deviation require fewer stages. This observation provides a justification to start the SMC algorithm from a tempered posterior of the approximating model rather than the full posterior.\footnote{This observation is related to the well-known importance sampling result that the proposal density should have fatter tails than the target density, .e.g., see \cite{Geweke1989}.}

In Panel (1,2) we depict the tempering schedules for $\mu = -1$ and various values of $\sigma$. Because the mean of the approximating density is different from the mean of the target density, increasing the standard deviation $\sigma$ from 0.2 to 2.0 increases the overlap of the two densities, decreases ${\cal D}(M_0,M_1)$, and leads to a steeper tempering schedule. The runtime pattern in Panel (2,1) mirrors the pattern of the average number of stages, because the runtime increases linearly in the number of stages. Finally, we show the standard deviation of the Monte Carlo approximation of $\mathbb{E}[\theta|Y,M_1]$ as a function of ${\cal D}(M_0,M_1)$ in Panel (2,2). Here no clear relationship with the discrepancy between approximate and target posterior emerges. Because the tempering schedule is chosen adaptively, the accuracy can be as good for large mismatches as it can be for small discrepancies, but it takes more time in the former case.

We deduce from this example that (i) the speed of the model tempering approach depends on the discrepancy between the approximating and the target density. (ii) Except for the additional runtime, the algorithm still works for fairly large discrepancies between the two densities. (iii) It might be desirable to start from a tempered rather than the full posterior of the approximating model.

\section{Illustration 2: A VAR with Stochastic Volatility}
\label{sec:VARSV}

This section demonstrates the benefits of model tempering in the context of a VAR with SV. The illustration is based on the VAR analysis in \cite{AruobaMlikotaSchorfheideVillalvazo2021}, except that we do not include a censored endogenous variable. The VAR model that is used as data generating process (DGP) and then estimated based on simulated data is presented in Section~\ref{subsec:VARSV.setup}. The parameterization of the VAR and the tuning of the SMC algorithm are summarized in Section~\ref{subsec:VARSV.parameters}. The numerical results are discussed in Section~\ref{subsec:VARSV.results}.

\subsection{VAR Specification}
\label{subsec:VARSV.setup}

Model $M_1$ is taken to be a bivariate VAR(1) with stochastic volatility:
\be
y_t = \Phi_1 y_{t-1} + \Phi_c + \mbox{chol}(\Sigma) \varepsilon_t , \quad \varepsilon_t \sim N(0,D_t) , \quad D_t = \mbox{diag}(d_t),
\label{eq:VARSV.M1}
\ee
where $\Sigma$ is a symmetric positive definite matrix and $\mbox{chol}(\cdot)$ is the lower-triangular Cholesky factor. Let $d_t = [d_{1,t},d_{2,t}]'$ and assume that its elements evolve according to
\be
\ln d_{it} = \rho_i \ln d_{it-1} + \xi_i \eta^i_t, \quad \eta^i_t \sim N(0,1), \quad i=1,2.
\label{eq:VARSV.M1.dt}
\ee
The presence of stochastic volatility renders this model nonlinear. However, the conditional linearity makes the likelihood evaluation relatively straightforward. We use a Bootstrap Particle Filter (BSPF) with $M_{bspf}=100$ particles, as outlined in the Online Appendix, to sequentially integrate out the latent volatility states. The BSPF likelihood evaluation can be conveniently integrated into the SMC sampler described in Algorithm~\ref{algo:smc}.\footnote{The use of a particle filter to evaluate the likelihood in the SMC posterior sampler results in a SMC$^2$ algorithm, as discussed in \cite{ChopinJacobPapas2012}.} Moreover, this computational strategy is very similar to a Bayesian estimation approach widely-used for the estimation of nonlinear DSGE models. We will use the BSPF also in Section~\ref{sec:DSGE}.

The approximating model $M_0$ is identical to $M_1$ except that it ignores stochastic volatility. It is given by the homoskedastic VAR
\be
y_t = \Phi_1 y_{t-1} + \Phi_c + u_t, \quad u_t \sim N(0,\Sigma).
\label{eq:VARSV.M0}
\ee
In other words, $M_0$ is obtained by setting $\xi_i = 0 \; \forall \; i$. This restriction renders $(\rho_i,\xi_i)$ non-identified. One obtains the standard analytical expression for the likelihood of a VAR, which as a result can be evaluated instantaneously. Thus, one important condition that makes model tempering attractive is satisfied: the evaluation of the likelihood function for the approximating model is considerably faster than the evaluation of the target model's likelihood.

We use a version of the Minnesota prior for $(\Phi_1,\Phi_c,\Sigma)$. The (marginal) prior for each $\rho_i$ is a Uniform distribution, while that for $\xi_i$ is an inverse Gamma distribution. Further details on the prior are provided in the Online Appendix. The prior specification is the same for all DGP parameterizations.

\subsection{Parameterization of DGP and Tuning of Algorithm}
\label{subsec:VARSV.parameters}

\begin{figure}[t!]
	\caption{Stochastic Volatility Paths and $M_0$ vs. $M_1$ Posteriors}
	\label{fig:VARSV.SV.posteriors}
	\begin{center}
		\begin{tabular}{ccc}
			DGP 1 & DGP 2 & DGP 3 \\
			\includegraphics[width=2.0in]{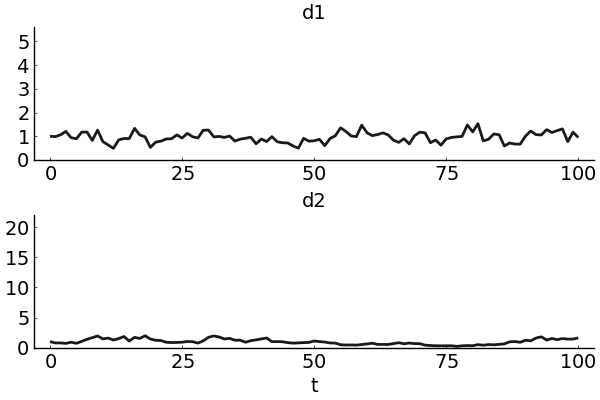} &
			\includegraphics[width=2.0in]{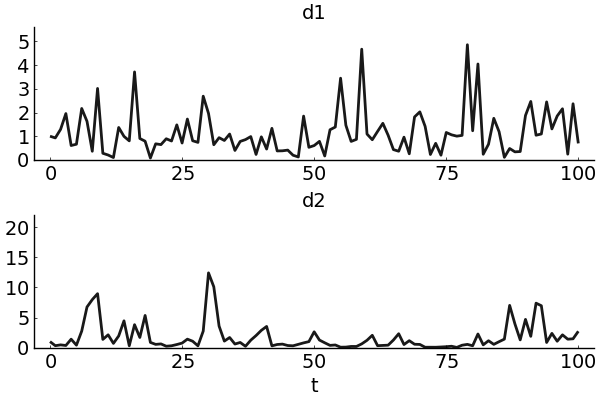} &
			\includegraphics[width=2.0in]{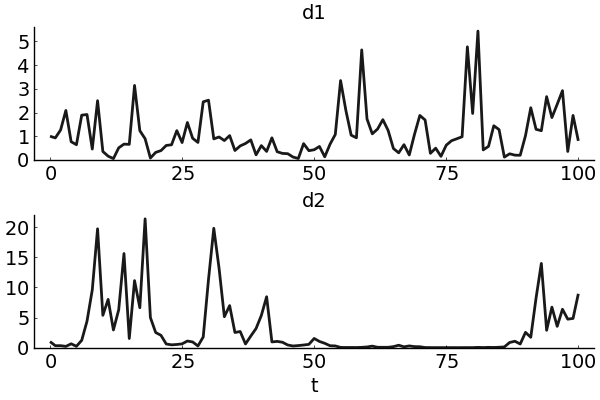}  \\
			\includegraphics[width=2.0in]{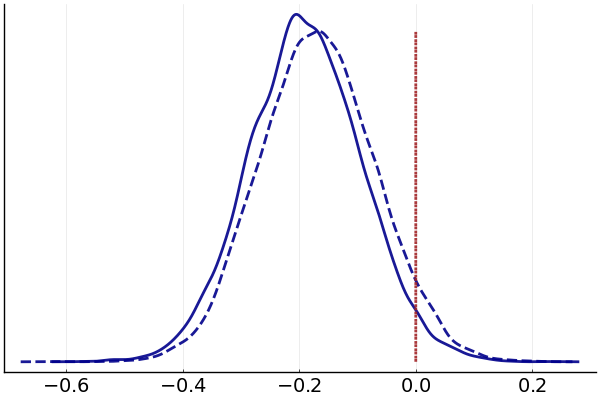} &
\includegraphics[width=2.0in]{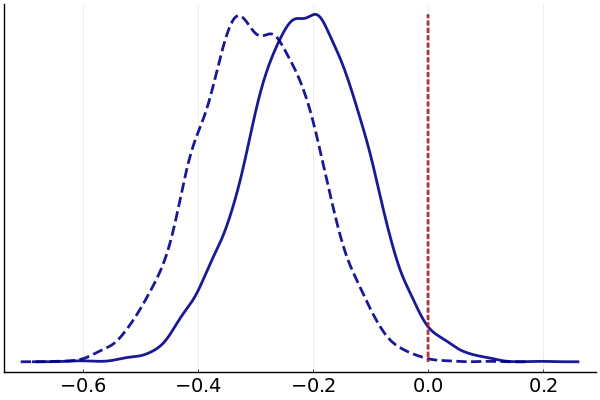} &
\includegraphics[width=2.0in]{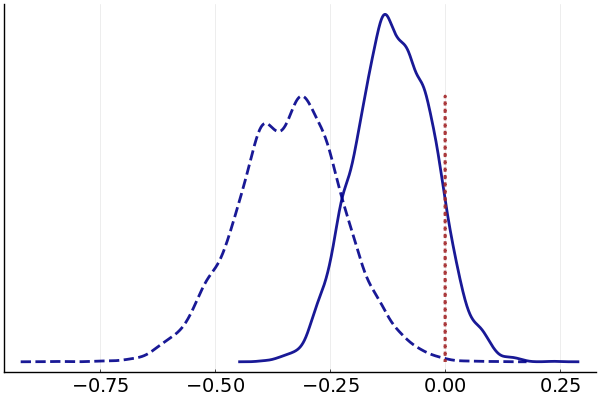}
		\end{tabular}
	\end{center}
	{\footnotesize {\em Notes:} Top row: simulated volatility paths $d_{1t}$ and $d_{2t}$. Bottom row:  $M_0$ (dashed black) versus $M_1$ (solid blue) posterior densities for $\Phi_{1,21}$. Dotted vertical line indicates true value. }\setlength{\baselineskip}{4mm}
\end{figure}

Estimation is conducted on $T=100$ observations simulated from model $M_1$. We use the following parameterization for $(\Phi_1,\Phi_c,\Sigma)$:
\[
	\Phi_1 =
    \begin{bmatrix} 0.6 & 0.3 \\ 0.0 & 0.4 \end{bmatrix}, \quad
    \Phi_c =
    \begin{bmatrix} 0.0  \\  0.0 \end{bmatrix}, \quad
    \Sigma =
    \begin{bmatrix} 1.0 & 0.0 \\ 0.7 & 1.0 \end{bmatrix} \cdot \begin{bmatrix} 1.0 & 0.7 \\ 0.0 & 1.0 \end{bmatrix}
    = \begin{bmatrix} 1.00 & 0.70 \\ 0.70 & 1.49 \end{bmatrix}.
\]
The closeness of the posteriors under the target model (VAR with SV) and the approximating model (homoskedastic VAR) depends on the parameterization of the stochastic volatility processes. We consider three different parameterizations which are summarized in Table~\ref{tab:dgp.sv}. Under DGP~1 (baseline) the standard deviations of the log volatility innovations are relatively small. This implies that the $\ln d_{it}$s only exhibit modest time variation and the homoskedastic specification provides a good approximation. Under DGP~2 the volatility innovations have larger standard deviations but the log volatility processes are less persistent, implying large yet short-lived swings in volatility. Finally, DGP~3 combines the baseline values for $\rho_i$ with the large values of $\xi_i$ also considered under DGP~2, implying the largest distance between the approximating model and the target model. This is confirmed in  Figure~\ref{fig:VARSV.SV.posteriors}. The panels in the top row show the volatility paths, $d_{1t}$ and $d_{2t}$, and the bottom row illustrates the resulting discrepancy between the $M_0$ and $M_1$ posteriors, using the parameter $\Phi_{1,21}$ as an example.


\begin{table}[t!]
	\caption{Parameterizations of the SV Processes}
	\label{tab:dgp.sv}
	\begin{center}
		\begin{tabular}{lcccc} \hline \hline
			   & $\rho_1$ & $\rho_2$ & $\xi_1$ & $\xi_2$ \\ \hline
		DGP 1  & 0.50     & 0.90     & 0.20    & 0.20  \\
		DGP 2  & 0.20     & 0.60     & 0.80    & 0.90  \\
		DGP 3  & 0.50     & 0.90     & 0.80    & 0.90 \\ \hline
		\end{tabular}
	\end{center}
\end{table}


\begin{figure}[t!]
	\caption{Approximate Distributions for $\Phi_{1,21}$, DGP 2}
	\label{fig:VARSV.approximate.distributions}
	\begin{center}
		\includegraphics[width=5in]{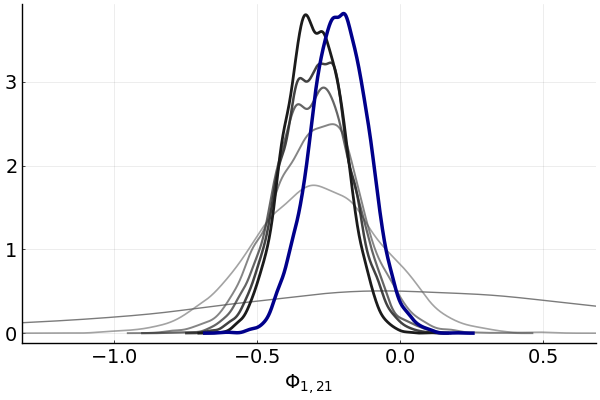}
	\end{center}
	{\footnotesize {\em Notes:} The approximating posterior densities obtained from the tempered $M_0$ likelihood function for $\psi_* \in \{0.0,0.2,0.4,0.6,0.8,1.0\}$ are plotted in shades (the larger $\psi_*$ the darker) of gray. The $M_1$ posterior is depicted in blue.}\setlength{\baselineskip}{4mm}
\end{figure}

We use an adaptive tempering schedule with $\alpha=0.95$, as described in Section~\ref{subsec:smc.implementation}, to ensure that the number of SMC stages and hence the computational time adjust endogenously to the distance between the proposal and the target density. We initialize the SMC algorithm based on the following set of tempered $M_0$ posteriors:
\[
\pi_0(\theta) \propto p(Y|\theta,M_0)^{\psi_*} p(\theta), \quad \psi_* \in \{0.0,0.2,0.4,0.6,0.8,1.0\},
\]
where $\psi_*=0$ corresponds to likelihood tempering, i.e., estimation of $M_1$ without using information from model $M_0$. Higher values for $\psi_*$ increasingly tilt the proposal density away from the prior distribution towards the posterior under the proxy model $M_0$. For $\phi_*=1$, the proposal density coincides with the posterior under $M_0$. This is illustrated in Figure~\ref{fig:VARSV.approximate.distributions} which shows along with the $M_1$ target posterior the sequence of approximating posterior distributions for $\Phi_{1,21}$ under DGP~2 obtained from the $\psi_*$-tempered $M_0$ likelihood function.

The number of particles in the SMC sampler is set to $N=500$. For each DGP and $\pi_0(\theta)$ we run the SMC algorithm $N_{run}=200$ times. We subsequently report averages across the $N_{run}$ runs and assess the variance of the Monte Carlo approximations across runs.

\subsection{Results}
\label{subsec:VARSV.results}

\begin{figure}[t!]
	\caption{Log MDD and Precision}
	\label{fig:VARSV.sameres}
	\begin{center}
    \begin{tabular}{cc}
    	Mean (DGP 1, $N=500$) & Standard Deviation \\
		\includegraphics[width=0.47\textwidth, clip]{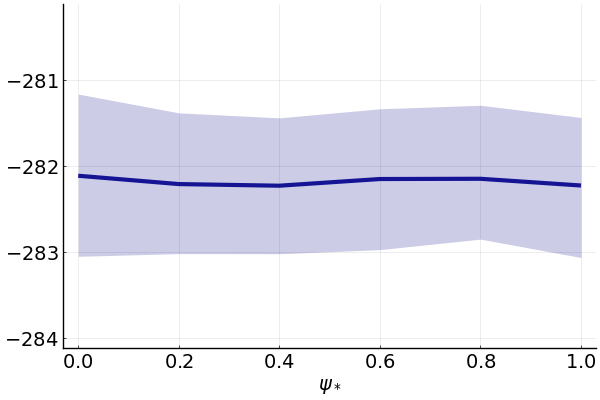} &
		\includegraphics[width=0.47\textwidth, clip]{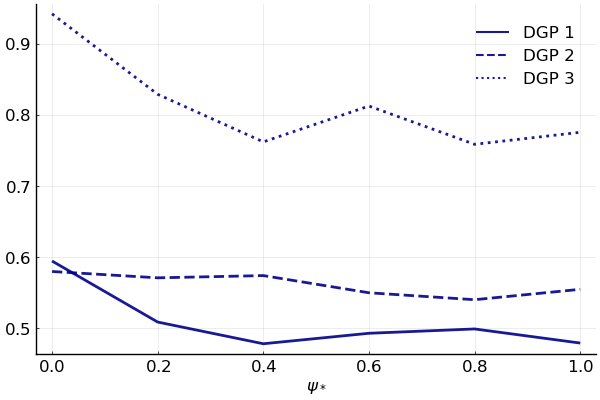}
    \end{tabular}
    \end{center}
	{\footnotesize {\em Notes:} The left panel depicts the mean and 90\% credible bands based on $N_{run}=200$ runs for DGP~1 with $N=500$ particles. The right panel shows the standard deviation of log MDD across the runs for all considered setups. }\setlength{\baselineskip}{4mm}
\end{figure}

The left panel of Figure~\ref{fig:VARSV.sameres} plots average (across multiple SMC runs) Monte Carlo approximates of the log MDD of model $M_1$, $\ln p(Y|M_1)$, under DGP~1 as a function of $\psi_*$, i.e. as a function of the degree of model tempering used in the construction of $\pi_0(\theta)$. The flat line confirms that the Monte Carlo approximations are the same regardless of $\pi_0(\theta)$, as we are numerically approximating the same object regardless of $\psi_*$. The shaded area is a 95\% credible band for the log MDD.
Figure~\ref{appfig:sameres_post} in the Online Appendix confirms that the Monte Carlo approximations for posterior mean, variance, 5th and 95th percentiles of the VAR parameters are also invariant to $\psi_*$. Moreover, the result holds not just under DGP~1, but also the other two DGPs (not shown in the figures).

The right panel of Figure~\ref{fig:VARSV.sameres} shows the standard deviation of the Monte Carlo approximation of the log MDD as a function of $\psi_*$ for the three different DGPs. For DGP~1 and DGP~3 the standard deviations are weakly decreasing in $\psi$. The biggest drop occurs between $\psi_*=0$ and $\psi_*=0.2$. For DGP~2 the profile is approximately flat, that is, based on the discrepancy between approximating density and target density, the algorithm adjusts the number of stages to keep the accuracy of the Monte Carlo approximation approximately constant.

\begin{figure}[t!]
	\caption{VAR-SV: Computational Times and Initial Variance of Particle Weights}
	\label{fig:VARSV.res_time}
	\begin{center}
		\begin{tabular}{cc}
            Relative Runtime & Importance Weight Variance \\
			\includegraphics[width=0.47\textwidth, clip]{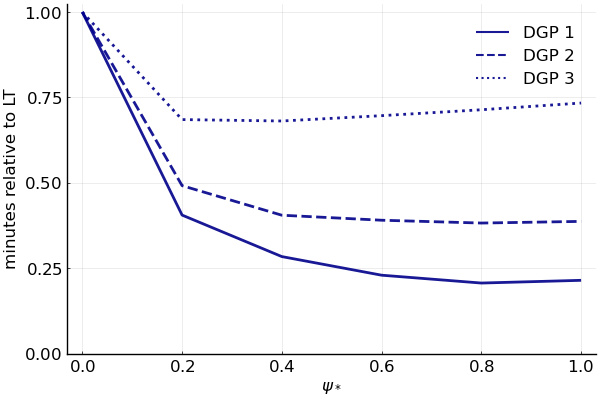} &
     		\includegraphics[width=0.47\textwidth, clip]{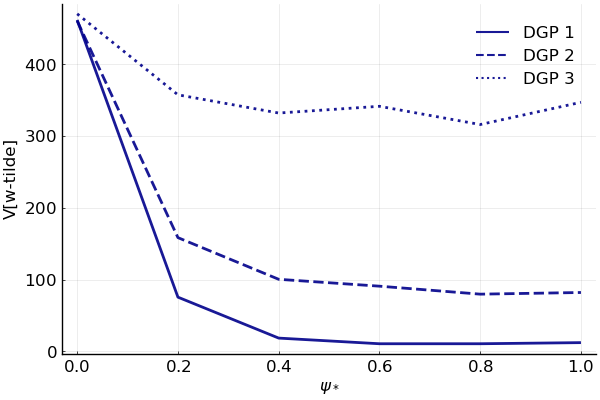}
		\end{tabular}
	\end{center}
	{\footnotesize {\em Notes:} The left panel plots the computational time relative to LT ($\phi_{N_\phi}(M_0)=0$) for the three DGPs (all with $N=500$). The right panel depicts the variance of particle weights $\tilde{W}^i(\psi_*)$ defined in (\ref{eq:defWpsistar}). In both panels we depict averages across the $N_{run}=200$ runs.}\setlength{\baselineskip}{4mm}
	\end{figure}

We now proceed by examining how $\psi_*$ affects the runtime of the SMC algorithm. The main result is presented in Figure~\ref{fig:VARSV.res_time}. The left panel compares the runtime profiles (normalized by the LT runtime) across the three DGPs. Incorporating information from model $M_0$ in the construction of the proposal density drastically reduces the runtime. The ratio of likelihood evaluation times for $M_0$ and $M_1$ is $\tau_0/\tau_1 = 9.14$. For DGP~1 the runtime monotonically decreases as the proposal is increasingly tilted towards the posterior of the approximate model $M_0$, with the largest reduction of close to 80\% obtained for $\psi_* \in \{0.6,0.8,1.0\}$. The runtime reduction is largest for DGP~1, followed by DGP~2, while DGP~3 is associated with the smallest reduction.

The steepest decrease in runtime occurs at $\psi_*=0.2$, which means that incorporating just a bit of information from the approximate model $M_0$ when constructing the proposal density for the posterior of model $M_1$ goes a long way in reducing the runtime. Adding more information helps little at best and might be even counterproductive, as is the case for DGP~3. In principle, for an even larger distance between the posteriors under the two models, it is conceivable that there is no runtime reduction at all. In this case, the second condition stated in Section~\ref{subsec:smc.tempering} would be violated.

The runtime benefits of model tempering decrease with the distance between the posteriors under the target model $M_1$ and the approximate model $M_0$.\footnote{The runtime for DGP~3 also exhibits the most variation across multiple runs; see Figure~\ref{appfig:runtime.schedule.all} in the Online Appendix. Depending on the run, there could be many or only very few particles in the small area to which both posteriors assign some positive probability mass.} In our VAR application, this distance increases with the nonlinearities generated by the SV specification, which are strongest for DGP~3. This distance is visualized in Figures~\ref{appfig:distEvol2D_1}, \ref{appfig:distEvol2D_2} and~\ref{appfig:distEvol2D_3} in the Online Appendix. While for DGP~1 all marginal posteriors align very well for the two models, for DGP~3 there are some parameters for which there is little overlap of probability mass between the two distributions.

We discussed in Section~\ref{subsec:smc.gains} that this distance could be assessed {\em ex ante}, without having completed the SMC run for $M_1$, by computing the variance  of the importance sampling weights defined in (\ref{eq:defWpsistar}). The variances as a function of $\psi_*$ for the three DGPs are depicted in the right panel of Figure~\ref{fig:VARSV.res_time}. The variance profiles look very similar to the runtime profiles in the left panel. For $\psi_*=0$ the variance is approximately equal to the number of particles minus one, $N-1$, which means that one of the particles has weight $N$ and the remaining particles have weight zero. For $\psi_*=0.2$ the variance is considerably lower: it is 75 for DGP~1 and 357 for DGP~3.

To summarize, in this VAR-SV application model tempering is able to reduce the relative runtime by between 27\% (DGP~3) and 79\% (DGP~1) and increase the precision of Monte Carlo approximations (DGP~1 and~3).



\section{Illustration 3: A Nonlinear DSGE Model}
\label{sec:DSGE}

Finally, we consider the estimation of a nonlinear DSGE model, which is computationally demanding for two reasons: first, the model needs to be solved and, second, the evaluation of the likelihood function requires a nonlinear filter. For the latter task, we will use a particle filter similar to the one used to estimate the VAR with SV in Section~\ref{sec:VARSV}. In our numerical illustration, we consider a real business cycle model with asymmetric quadratic capital adjustment costs. The adjustment cost parameters let us control the degree of nonlinearity. The model economy is described in Section~\ref{subsec:DSGE.setup}, the configuration of the simulation experiment is summarized in Section~\ref{subsec:DSGE.parameters}, and the simulation results are presented in Section~\ref{subsec:DSGE.results}.

\subsection{Model Specification}
\label{subsec:DSGE.setup}

The model economy consists of a representative household and a representative firm. The household consumes $C$, supplies labor in the amount of $L$, and owns the capital stock $K$. The firm hires labor and rents capital to produce a single good that can be used for consumption and investment. The model dynamics can be described as the solution to the following social planner problem:
\begin{eqnarray}
	V(K,S) &=& \underset{C,L,K'}{\max} \; \frac{C^{1-\tau}-1}{1-\tau} - B \frac{L^{1+1/\nu}}{1+1/\nu} + \beta \mathbb{E}_{S'|S} [V(K',S')]  \\
	&  \text{s.t.} &  C + I + K \Phi(K'/K) = Y, \nonumber \\
	& & Y =  Z K^\alpha L^{1-\alpha}, \nonumber \\ & & I = K' - (1-\delta)K. \nonumber
\end{eqnarray}
Households derive utility from consumption and disutility from labor. The parameter $\beta$ is the discount factor, $\tau$ determines the risk aversion, and $\nu$ is the Frisch labor supply elasticity. The parameter $\alpha$ is the capital share parameter, and $\delta$ the depreciation rate. Total factor productivity $Z$ and the preference process $B$ evolve exogenously according AR(1) laws of motion:
\begin{eqnarray}
	Z &= Z_*e^{\hat z}, \quad \hat z' = \rho_z \hat z + \sigma_z \varepsilon_z' , \\
	B &= B_*e^{\hat b}, \quad \hat b' = \rho_b \hat b + \sigma_b \varepsilon_b' \; . \nonumber
\end{eqnarray}
Thus, $\varepsilon_z'$ can be thought of as a supply and $\varepsilon_x'$ as a demand shock. Here we adopt the convention that for a variable $X$, $X_*$ denotes the steady state and $\hat{x}$ denote log deviations from the steady state.

For the adjustment cost function we use a linex function which is asymmetric:
\be
   \Phi(K'/K) = \phi_1 \left( \frac{\exp(-\phi_2(K'/K-1))+ \phi_2(K'/K-1)-1}{\phi_2^2} \right).
\ee
The parameter $\phi_1$ controls the overall level of adjustment costs and $\phi_2$ determines the asymmetry. Notice that as $\phi_2 \longrightarrow 0$ the adjustment costs become quadratic around the replacement investment level at which $K'/K=1$. If $\phi_2 > 0$, then it is more costly to reduce the capital stock than it is to augment the capital stock.

Model $M_1$ refers to a nonlinear solution of the RBC growth model, obtained using a second-order perturbation around the steady state. The approximate model $M_0$ is obtained by conducting a first-order linearization. The models are estimated based on observations for output, investment, and hours worked. We denote the observed variables by an o-superscript. The measurement equations, now with $t$ subscripts, take the form:
\begin{eqnarray}
\ln Y_t^o &=& \ln Y_t + \eta_{Y,t}, \quad \eta_{Y,t} \sim N(0,\sigma^2_Y), \label{eq:DSGE.ME} \\
\ln I_t^o &=& \ln I_t + \eta_{I,t}, \quad \eta_{I,t} \sim N(0,\sigma^2_I),  \nonumber  \\
\ln L_t^o &=& \ln L_t + \eta_{L,t}, \quad \eta_{L,t} \sim N(0,\sigma^2_L), \nonumber
\end{eqnarray}
where $\ln Y_t$, $\ln I_t$, and $\ln L_t$ are the model-implied series and the $\eta_t$s are measurement errors.
The measurement errors facilitate the use of a particle filter in combination with the nonlinear DSGE model solution. Moreover, they help to overcome the singularity problem generated by fitting a DSGE model with two shocks to three observables. We include the measurement errors in both data generation and estimation and fix their standard deviations such that the variance of the measurement error is approximately 5\% of the variation of the series $\ln Y_t$, $\ln I_t$, and $\ln L_t$, respectively.\footnote{The values that we use are $\sigma_Y=.006$, $\sigma_I = .004$, and $\sigma_L = .004$.}

\subsection{Model Parameterization and Tuning of Algorithm}
\label{subsec:DSGE.parameters}

To facilitate the estimation, we reparameterize the model as follows. First, we express the discount factor $\beta$ as a function of an annualized real interest rate (in percentages) $r = 400(1/\beta-1)$. Second, instead of parameterizing the model in terms of steady states of the exogenous processes $(Z_*,B_*)$, we use the steady states of output and labor, $(Y_*,L_*)$, which we set equal to one for both data generation and estimation. Because our observations $\ln Y_t^o$, $\ln I_t^o$, and $\ln L_t^o$ do not contain direct information on the steady state interest rate and the amount of investment necessary to replace depreciating capital stock, we fix $r$ and $\delta$ at their true values. We collect the parameters that are being estimated in the vector $\theta$:
\[
\theta = [\tau, \nu,\alpha,\phi_1,\phi_2,\rho_z,\rho_b,100 \sigma_z,100 \sigma_b]'.
\]

\begin{table}[t!]
	\caption{DGP and Prior}
	\label{tab:DSGE.dgpprior}
	\begin{center}
		\begin{tabular}{lllllrr} \hline\hline
			&	True &     \multicolumn{5}{c}{ Prior Distribution }   \\
			& Value & Density   &  P(1)  &  P(2)  & HPD Low   &   HPD High   \\
			\hline
			$r$           &  3.00     & \multicolumn{5}{c}{fixed at 3.00} \\
			$\delta$      &  0.08     & \multicolumn{5}{c}{fixed at 0.08} \\ \hline
			$\tau$        &  2.00     & $\mathcal{G}$ & 1.00 & 1.00 & .0005 & 2.27\\
			$\nu$         &  1.00     & $\mathcal{G}$ & 0.50 & 0.30 & 0.07 & 0.87 \\
			$\alpha$      &  0.35     & $\mathcal{B}$ & 0.35 & 0.05 & 0.27 & 0.43 \\
			$\phi_1$ 			& 50.0  		& $\mathcal{G}$ & 30.0 & 15.0 & 8.47 & 50.8\\
			$\phi_2$ 			& 200 			& $\mathcal{N}$ & 0 	 & 75.0 & -123 & 116\\
			$\rho_z$      &  0.95     & $\mathcal{B}$ & 0.6  & 0.15 & 0.35 & 0.80\\
			$\rho_b$      &  0.90     & $\mathcal{B}$ & 0.6  & 0.15 & 0.38 & 0.82\\
			$100 \sigma_z$    &  2.00 & $\mathcal{IG}$ & 1.50 & 5.00 & 0.45 & 2.27\\
			$100 \sigma_b$    &  1.60 & $\mathcal{IG}$ & 1.50 & 5.00 & 0.53 & 2.35\\
			\hline
		\end{tabular}
	\end{center}
	{\footnotesize {\em Notes:} We set $Y_*=L_*=1$ and we define $r=400(1/\beta-1)$. $\mathcal{G} $  is Gamma distribution; $\mathcal{B}$ is Beta distribution; $\mathcal{IG} $  is Inverse Gamma distribution; and ${\cal N}$ is Normal distribution, and ${\cal U}$ is Uniform distribution. P(1) and P(2) are mean and standard deviations for ${\cal B}$, ${\cal G}$, and ${\cal N}$ distributions. The ${\cal U}$ distribution is parameterized in terms of lower and upper bound. The ${\cal IG}$ distribution is parameterized as scaled inverse $\chi^2$ distribution with density $p(\sigma^2|s^2,\nu) \propto (\sigma^2)^{-\nu/2-1} \exp[-\nu s^2/(2 \sigma^2)]$, where P(1) is $\sqrt{s^2}$ and P(2) is $\nu$. The density of $\sigma$ is obtained by the change of variables $\sigma = \sqrt{\sigma^2}$. HPD(Low,High) refers to the boundaries of 90\% highest prior density intervals. }\setlength{\baselineskip}{4mm}
\end{table}

As in Section~\ref{sec:VARSV}, the estimation is conducted using data simulated from model $M_1$. The parameterization of the DGP is summarized in Table~\ref{tab:DSGE.dgpprior}. Most of the parameter values are similar to values commonly found in the DSGE model literature, except that we scale up the shock standard deviations and use fairly large asymmetric adjustment costs by setting $\phi_1 = 50$ and $\phi_2 = 200$.  We plot simulated sample paths in the Online Appendix. The length of the estimation sample is $T=80$. The remaining columns of Table~\ref{tab:DSGE.dgpprior} describe the prior distribution for the Bayesian estimation.

In the SMC algorithm we use $N=1,000$ particles to represent the distribution of $\theta$, $N_{MH}=2$ Metropolis-Hastings steps in the mutation, and an adaptive tempering schedule with $\alpha=0.95$. While the likelihood function associated with $M_0$ can be evaluated with the Kalman filter, a nonlinear filter is required to compute the likelihood function of $M_1$. We use the same BSPF that was used in Section~\ref{sec:VARSV} for the VAR estimation with $M_{bspf}=2,000$ particles.

\subsection{Results}
\label{subsec:DSGE.results}

\begin{figure}[t!]
	\caption{Results from the RBC Model}
	\label{fig:RBC.results}
	\begin{center}
		\begin{tabular}{cc}
			(1,1) Log MDD & (1,2) Tempering Schedules \\
			\includegraphics[width=0.47\textwidth, clip]{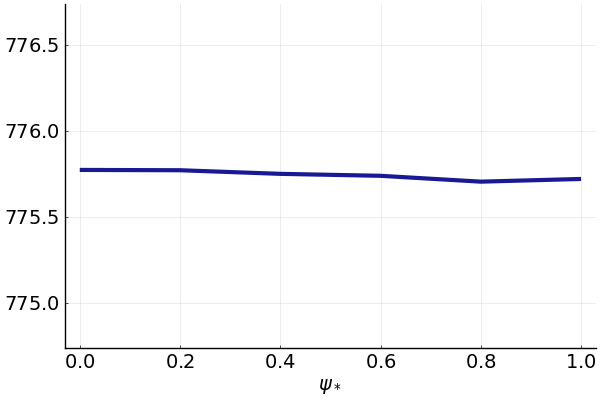} &
			\includegraphics[width=0.47\textwidth, clip]{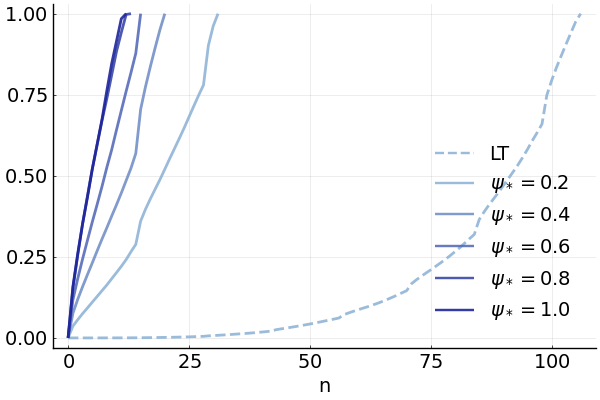} \\
			(2,1) Relative Runtime & (2,2) Importance Weight Variance \\
			\includegraphics[width=0.47\textwidth, clip]{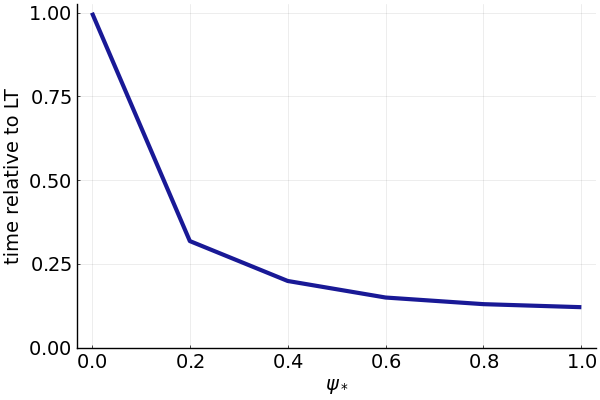} &
			\includegraphics[width=0.47\textwidth, clip]{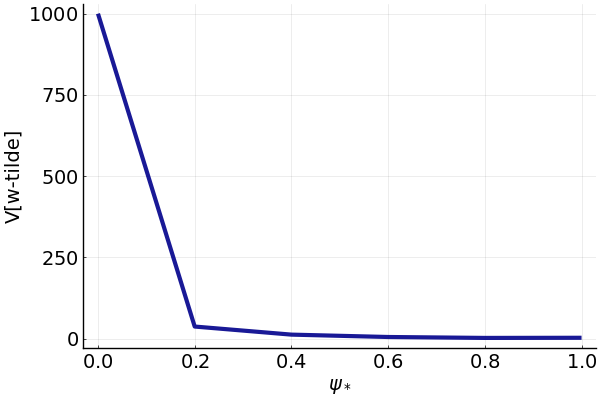}
		\end{tabular}
	\end{center}
	{\footnotesize {\em Notes:} Single run ($N_{run}=1$) }\setlength{\baselineskip}{4mm}
\end{figure}

As before, we consider $\psi_* \in \{0.0, 0.2, 0.4, 0.6, 0.8, 1.0\}$, where $\psi_* = 0$ corresponds to $M_1$ likelihood tempering. The relative time it takes to evaluate the $M_0$ and $M_1$ likelihood functions is $\tau_0/\tau_1 \approx 1/ 109$. This ratio depends on the number of particles $M_{bspf}$ used in the BSPF. Doubling $M_{bspf}$ would approximately double $\tau_1$ and cut the ratio in half. The numerical results from a single $N_{run}=1$ run of the model tempering SMC algorithm for the various values of $\psi_*$ are presented in Figure~\ref{fig:RBC.results}. The top left (1,1) panel shows the log MDD approximation, which is approximately constant as a function of $\psi_*$. This plot confirms that regardless of $\psi_*$ the SMC algorithm delivers the same approximations of the posterior distribution.

The tempering schedules are plotted in Panel (1,2). Starting from a (tempered) $M_0$ posterior drastically reduces the number of stages needed to reach the target posterior. This is consistent with the information provided by the importance weight variance in Panel (2,2). Reweighting draws from the prior $p(\theta)$ ($\psi_*=0$) to target the $M_1$ posterior would lead to a degenerate distribution of weights, whereby the weight of one draw is equal to one and all other weights are equal to zero. Starting with draws from the $\psi_*=0.2$ tempered $M_0$ posterior reduces the importance weight variance from $N-1=999$ to 37. Raising $\psi_*$ toward one, lowers the variance further.

Panel (2,1) depicts the runtime of the model-tempered SMC relative to the $M_1$ likelihood-tempered SMC. The biggest drop, from 1.0 to 0.3 occurs by raising $\psi_*$ from 0.0 to 0.2. Subsequent gains are smaller and the curve essentially turns flat from 0.8 onwards, converging to 0.12. In terms of absolute runtimes, on a Windows workstation with Intel(R) Xeon(R) CPU E5-2687 at 3.10GHz using 8 out of 10 cores model tempering with $\psi_*=1$ reduces the runtime of our JULIA code from 655 to 80 minutes. In the Online Appendix we plot the target and approximate (marginal) posterior densities for the DSGE model parameters. Despite the large $\phi_1$ and $\phi_2$ values, the nonlinearity generated by the DSGE model is not particularly strong and $M_0$ and $M_1$ posterior distributions are quite similar. Thus, when starting from $\psi_*=0.8$ or $\psi_*=1$ only minimal adjustments are required to turn the $M_0$ posterior draws into $M_1$ posterior draws which leads to substantial computational gains.


\section{Conclusion}
\label{sec:conclusion}

The implementation of posterior samplers for Bayesian inference often requires the explicit evaluation of likelihood functions. Likelihood calculations for macroeconometric models can be computationally demanding, because it may take a long time to solve the underlying structural model or it may be time-consuming to integrate out latent state variables. In this paper we documented how an SMC algorithm with model tempering can speed up posterior sampling for a VAR with stochastic volatility and a nonlinear DSGE model. The method is suitable for applications in which the likelihood evaluation for the target model is computationally costly and there is an approximating model for which the likelihood evaluation is fast and that generates a posterior that is not too different from the posterior of the target model.


\bibliography{MT_ref}


\clearpage

\renewcommand{\thepage}{A.\arabic{page}}
\setcounter{page}{1}

\begin{appendix}
	\markright{Online Appendix -- This Version: \today }
	\renewcommand{\theequation}{A.\arabic{equation}}
	\setcounter{equation}{0}

	\renewcommand*\thetable{A-\arabic{table}}
	\setcounter{table}{0}
	\renewcommand*\thefigure{A-\arabic{figure}}
	\setcounter{figure}{0}

	\begin{center}

		{\large {\bf Online Appendix: Sequential Monte Carlo with Model Tempering}}

		{\bf Marko Mlikota and Frank Schorfheide}
	\end{center}

\noindent This Appendix consists of the following sections:

\begin{itemize}
	\item[A.] Computational Details
	\item[B.] Illustration 2: VAR with Stochastic Volatility
	\item[C.] Illustration 3: A Nonlinear DSGE Model
\end{itemize}

\section{Computational Details}
\label{appsec:compute}

The presentations of the mutation algorithm in Section~\ref{appsubsec:compute.mutation} and the BSPF in in Section~\ref{appsubsec:compute.pf} are based on \cite{HerbstSchorfheide2016}.

\subsection{SMC Particle Mutation}
\label{appsubsec:compute.mutation}

\begin{algo}[Particle Mutation] \label{algo:mutation}\hspace*{1cm}\\
	In Step~2(c) in iteration $n$ of Algorithm~\ref{algo:smc}:
	\begin{enumerate}
		\item Compute an importance sampling
		approximation $\tilde{\Sigma}_n$ of $\mathbb{V}_{\pi_n}[\theta]$
		based on the particles $\{\theta_{n-1}^i,\tilde{W}_n^i \}_{i=1}^N$.
		\item Compute the average empirical rejection rate
		$\hat{R}_{n-1}(\hat{\zeta}_{n-1})$, based on the Mutation step in
		iteration $n-1$.  The average is computed across the $N_{blocks}$ blocks.
		\item Let $\hat{c}_1 = c^*$ and for $n>2$ adjust the scaling factor according to
		\[
		\hat{c}_{n} = \hat{c}_{n-1} f \big( 1-\hat{R}_{n-1}(\hat{\zeta}_{n-1}) \big),
		\]
		where
		\[
		f(x) = 0.95 + 0.10 \frac{e^{16(x - 0.25)}}{1 + e^{16(x - 0.25)}}.
		\]
		\item Define
		$\hat{\zeta}_n = \big[ \hat{c}_{n}, vech(\tilde{\Sigma}_n)' \big]'$.
		\item For each particle $i$, run $N_{MH}$ steps of a Random Walk Metropolis-Hastings Algorithm using the proposal density
		\be
		\vartheta^{i,m}_{n} |\hat{\zeta}_n \sim N \bigg( \theta^{i,m-1}_{n}, \hat{c}_n^2 \tilde{\Sigma}_{n} \bigg).
		\ee
	\end{enumerate}
\end{algo}

\subsection{(Particle) Filtering}
\label{appsubsec:compute.pf}

We use a bootstrap particle filter (BSPF) to approximate the likelihood function in the model with stochastic volatility. In the description of the filter we denote the latent state by $s_t$.

\begin{algo}[Bootstrap Particle Filter]
	\label{algo:bspf}
	\hspace*{1cm}\\[-3ex]
	\begin{enumerate}
		\item {\bf Initialization.} Draw the initial particles from the distribution $s_0^j \stackrel{iid}{\sim} p(s_0|\theta)$
		and set $W_0^j=1$, $j=1,\ldots,M$.

		\item {\bf Recursion.} For $t=1,\ldots,T$:
		\begin{enumerate}
			\item {\bf Forecasting $s_t$.} Draw $\tilde{s}_t^j$ from the state-transition density $p(\tilde{s}_t|s_{t-1}^j,\theta)$.
			\item {\bf Forecasting $y_t$.} Define the incremental weights
			\be
			\tilde{w}^j_t = p(y_t|\tilde{s}^j_t,Y_{1:t-1},\theta)
			\label{eq:copf.generalpfincrweight}
			\ee
			The predictive density $p(y_t|Y_{1:t-1},\theta)$
			can be approximated by
			\be
			\hat{p}(y_t|Y_{1:t-1},\theta) = \frac{1}{M} \sum_{j=1}^M \tilde{w}^j_t W_{t-1}^j.
			\label{eq:copf.generalpfpreddens}
			\ee
			\item Define the normalized weights
			\be
			\tilde{W}^j_t = \tilde{w}^j_t W^j_{t-1} \bigg/ \frac{1}{M} \sum_{j=1}^M \tilde{w}^j_t W^j_{t-1}.
			\ee
			\item {\bf Selection.} Resample the particles, for instance, via
			multinomial resampling. Let $\{ s_t^j \}_{j=1}^M$ denote $M$ iid draws from
			a multinomial distribution characterized by support points and weights
			$\{ \tilde{s}_t^j,\tilde{W}_t^j \}$ and set $W_t^j=1$ for $j=,1\ldots,M$.
			An approximation of $\mathbb{E}[h(s_t)|Y_{1:t},\theta]$ is given by $\bar{h}_{t,M} = \frac{1}{M} \sum_{j=1}^M h(s_t^j)W_{t}^j$.
		\end{enumerate}

		\item {\bf Likelihood Approximation.} The approximation of the log-likelihood function
		is given by \index{likelihood function}
		\be
		\ln \hat{p}(Y_{1:T}|\theta) = \sum_{t=1}^T \ln \left( \frac{1}{M} \sum_{j=1}^M \tilde{w}^j_t W_{t-1}^j \right).
		\ee
	\end{enumerate}
\end{algo}

\section{Illustration 2: A VAR with Stochastic Volatility}
\label{appsec:var}

\subsection{Prior Specification}
\label{appsubsec:var.prior}

\noindent {\bf Prior for $(\Phi_1,\Phi_2,\Sigma)$.} We use a Minnesota-type prior for the reduced-form VAR coefficients that appear in the homoskedastic version of the VAR in (\ref{eq:VARSV.M0}). The specification of the Minnesota prior follows \cite{DelNegroSchorfheide2012}. The prior is indexed by hyperparameters $\lambda_1$, $\lambda_2$, and $\lambda_3$, and is implemented through dummy observations stacked into $(Y^*,X^*)$. We use three sets of dummy observations, written as $Y^*_j = X^*_j \Phi + U_j$:
\begin{eqnarray*}
\begin{bmatrix}
\lambda_1 \underbar{s}_1  & 0 \\ 0 & \lambda_1 \underbar{s}_2 \\
\end{bmatrix}
&=& \begin{bmatrix}
\lambda_1 \underbar{s}_1  & 0 & 0 \\ 0 & \lambda_1 \underbar{s}_2 & 0\\
\end{bmatrix} \Phi + \begin{bmatrix}
u_{11} & u_{12} \\ u_{21} & u_{22}  \\
\end{bmatrix} \; , \\
\begin{bmatrix}
\lambda_2 \underbar{y}_1  & \lambda_2 \underbar{y}_2 \\
\end{bmatrix}
&=& \begin{bmatrix}
\lambda_2 \underbar{y}_1  & \lambda_2 \underbar{y}_2 & \lambda_2\\
\end{bmatrix} \Phi + \begin{bmatrix}
u_{11} & u_{12} \\ u_{21} & u_{22}  \\
\end{bmatrix} \; ,\\
\begin{bmatrix}
\underbar{s}_1  & 0 \\ 0 & \underbar{s}_2 \\
\end{bmatrix}
&=& \begin{bmatrix}
0  & 0 & 0 \\ 0 & 0 & 0\\
\end{bmatrix} \Phi + \begin{bmatrix}
u_{11} & u_{12} \\ u_{21} & u_{22}  \\
\end{bmatrix} \; ,
\end{eqnarray*}
where $\underbar{y}_i$ and $\underbar{s}_i$ are the mean and standard deviation of $y_i$. The first set of dummy observations implies that the VAR coefficients are centered at univariate unit-root representations. The second set of dummy observations implies that if the lagged value $y_{t-1}$ take the value $\underline{y}$, then the current value $y_t$ will be close to $\underline{y}$. The third set of dummy observations induces a prior for the covariance matrix of $u_t$ and is repeated $\lambda_3$ times. The dummy observations induce a conjugate MNIW prior for $(\Phi,\Sigma)$:
\[
\Sigma \sim IW(\underline{S},\underline{\nu}) \; , \quad \Phi | \Sigma \sim MN(\underline{\mu}, \Sigma \otimes \underline{P}^{-1}) \; ,
\]
with
\[
\underline{\nu} = T^* -k\; , \quad \underline{S} =S^* \; , \quad  \underline{\mu} = \Phi^* \; , \quad  \underline{P} = {X^*}'X^*\; ,
\]
where $\Phi^* = ({X^*}'X^*)^{-1} {X^*}'Y^*$ and $S^* = (Y^* - X^*\Phi^*)'(Y^* - X^*\Phi^*) $. We set $\lambda_1 = 1$,  $\lambda_2 = 1$, and $\lambda_3 = 3$.

\noindent {\bf Prior for $\rho_i$.} The prior for each $\rho_i$ is Uniform on $[0,1]$.


\noindent {\bf Prior for $\xi_i$.} The prior of $\xi_i$ is specified as an inverse Gamma distribution. It is parameterized as scaled inverse $\chi^2$ distribution with density $p(\xi^2|s^2,\nu) \propto (\xi^2)^{-\nu/2-1} \exp[-\nu s^2/(2 \xi^2)]$, where $\sqrt{s^2}$ is 0.3 and $\nu$ is 2.0. The density of $\xi_i$ is obtained by the change of variables $\xi = \sqrt{\xi^2}$.

\newpage
\subsection{Further Results for the VAR-SV}
\label{appsubsec.var.further.results}

\begin{figure}[h!]
	\caption{VAR-SV: Target and Approximate Posterior Densities for DGP 1}
	\label{appfig:distEvol2D_1}
	\begin{center}
		\begin{tabular}{ccc}
			\includegraphics[width=0.31\textwidth]{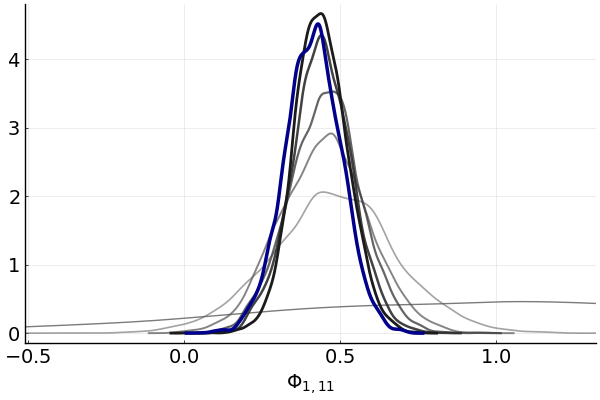}	&
			\includegraphics[width=0.31\textwidth]{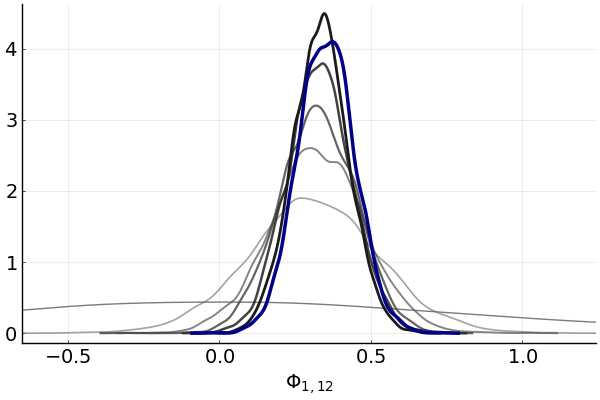} &
			\includegraphics[width=0.31\textwidth]{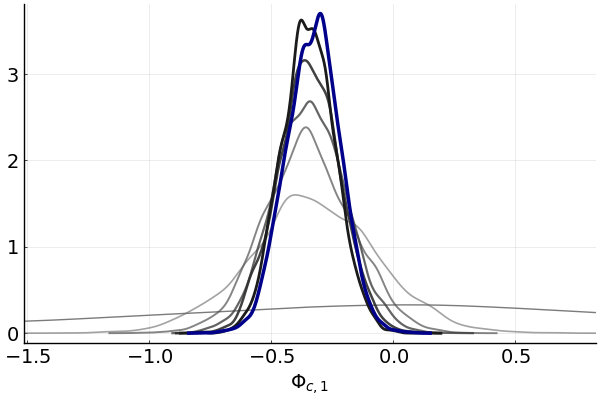} \\
			\includegraphics[width=0.31\textwidth]{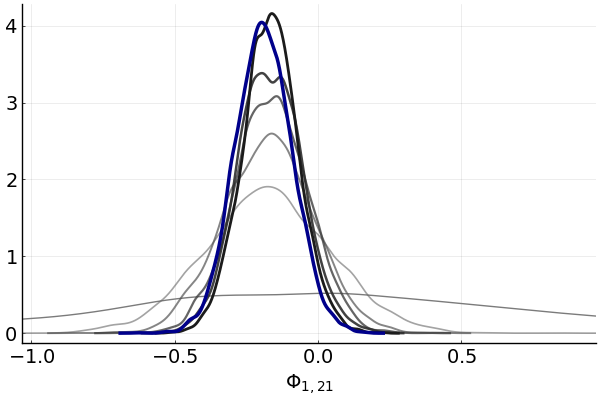}	&
	        \includegraphics[width=0.31\textwidth]{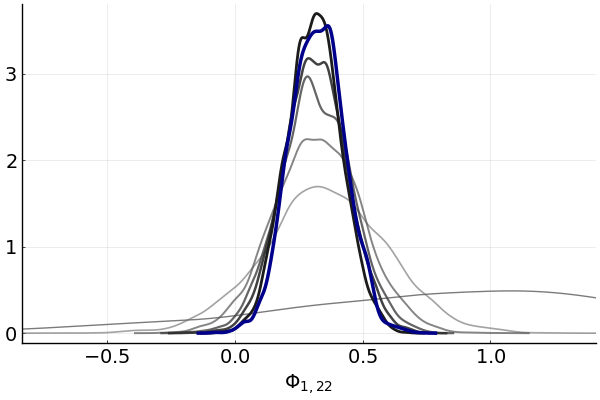} &
         	\includegraphics[width=0.31\textwidth]{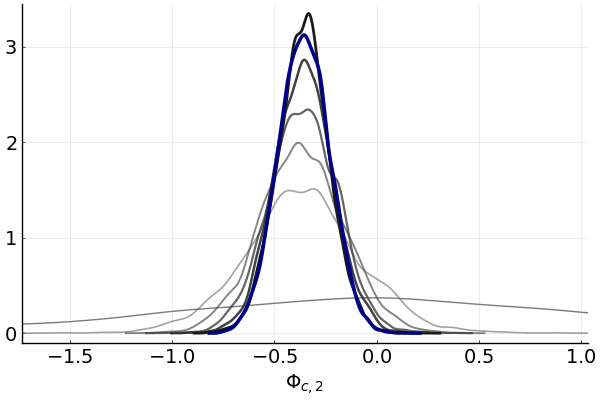} \\
			\includegraphics[width=0.31\textwidth]{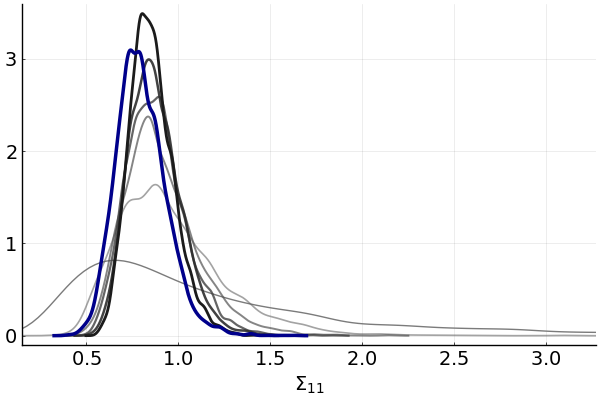}	&
            \includegraphics[width=0.31\textwidth]{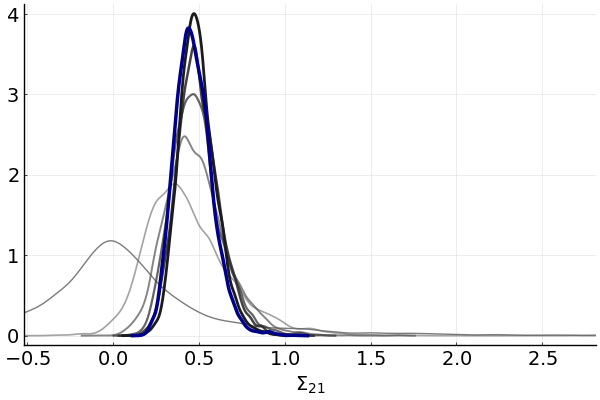} &
           \includegraphics[width=0.31\textwidth]{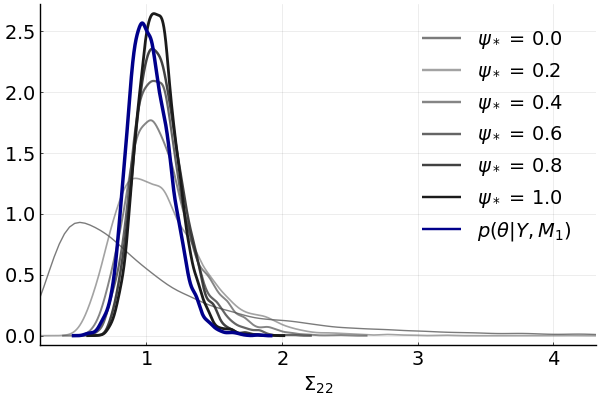} \\
		\end{tabular}
	\end{center}
	{\footnotesize {\em Notes:} Each plot refers to a different parameter. The approximating posterior densities obtained from the tempered $M_0$ likelihood function for $\psi_* \in \{0.0,0.2,0.4,0.6,0.8,1.0\}$ are plotted in shades (the larger $\psi_*$ the darker) of gray. The $M_1$ posterior is depicted in blue. The stochastic volatility parameters $\rho_i, \xi_i, \; i = 1,2$ are not displayed because model $M_0$ is uninformative for them.}\setlength{\baselineskip}{4mm}
\end{figure}

\begin{figure}[t!]
	\caption{VAR-SV: Target and Approximate Posterior Densities for DGP 2}
	\label{appfig:distEvol2D_2}
	\begin{center}
		\begin{tabular}{ccc}
			\includegraphics[width=0.31\textwidth]{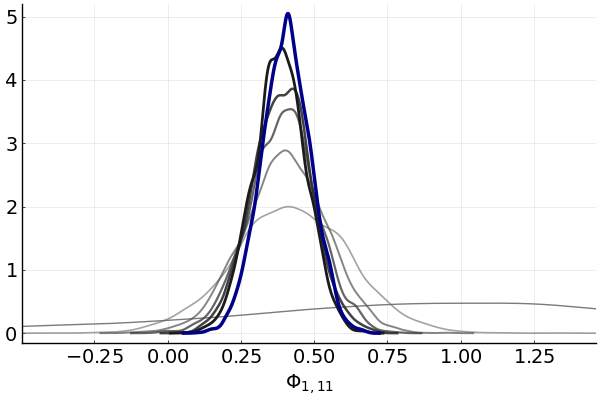}	&
			\includegraphics[width=0.31\textwidth]{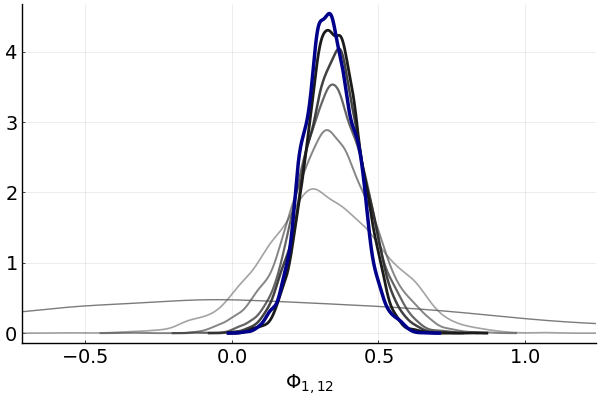} &
			\includegraphics[width=0.31\textwidth]{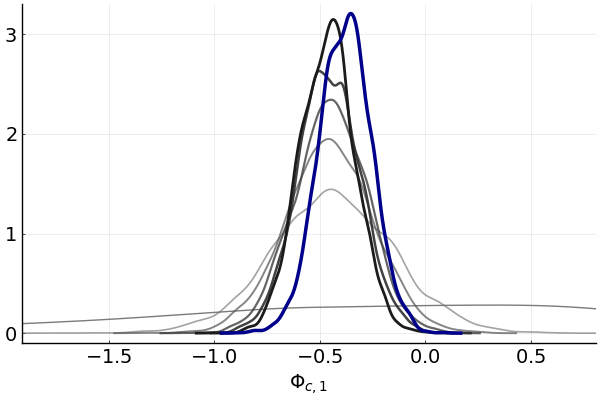} \\
			\includegraphics[width=0.31\textwidth]{figures/VAR-SV/dgp2_mSV_prSV1_mc1/plot_distEvol2D_p4.png}	&
			\includegraphics[width=0.31\textwidth]{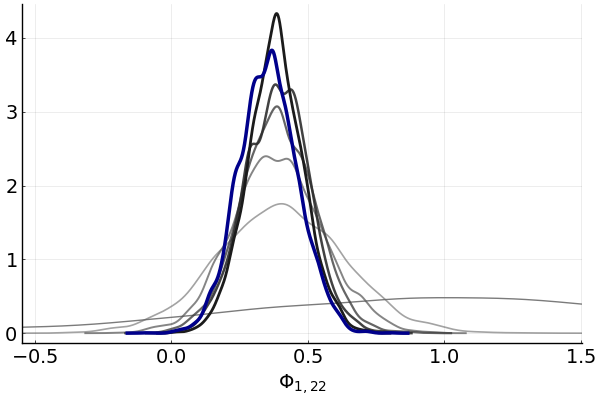} &
			\includegraphics[width=0.31\textwidth]{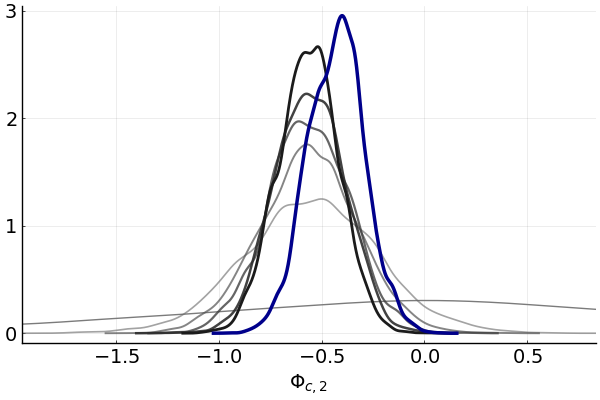} \\
			\includegraphics[width=0.31\textwidth]{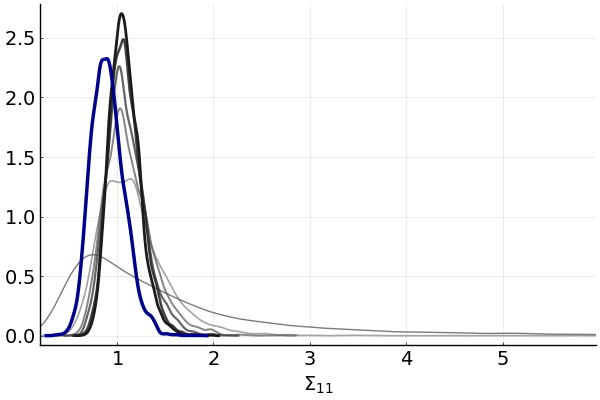}	&
			\includegraphics[width=0.31\textwidth]{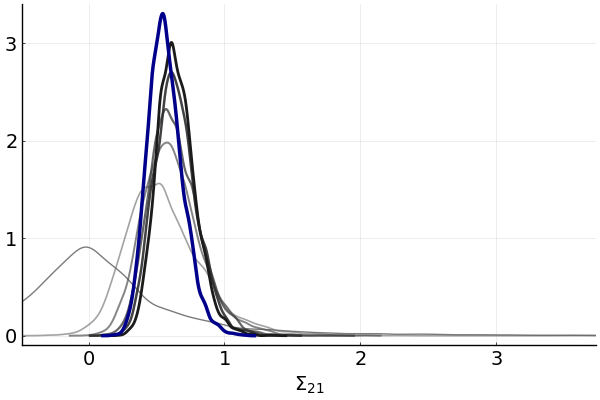} &
			\includegraphics[width=0.31\textwidth]{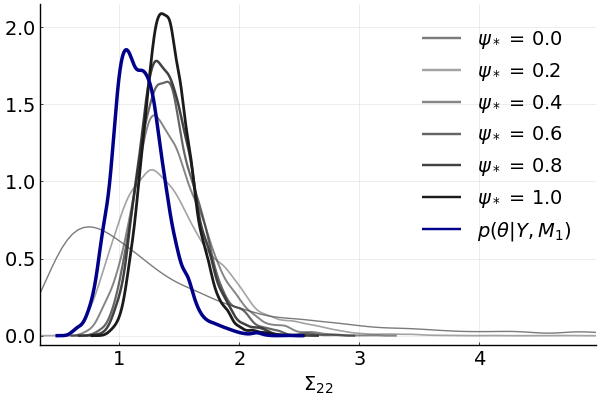} \\
		\end{tabular}
	\end{center}
	{\footnotesize {\em Notes:} Each plot refers to a different parameter. The approximating posterior densities obtained from the tempered $M_0$ likelihood function for $\psi_* \in \{0.0,0.2,0.4,0.6,0.8,1.0\}$ are plotted in shades (the larger $\psi_*$ the darker) of gray. The $M_1$ posterior is depicted in blue. The stochastic volatility parameters $\rho_i, \xi_i, \; i = 1,2$ are not displayed because model $M_0$ is uninformative for them.}\setlength{\baselineskip}{4mm}
\end{figure}

\begin{figure}[t!]
	\caption{VAR-SV: Target and Approximate Posterior Densities for DGP 3}
	\label{appfig:distEvol2D_3}
	\begin{center}
		\begin{tabular}{ccc}
			\includegraphics[width=0.31\textwidth]{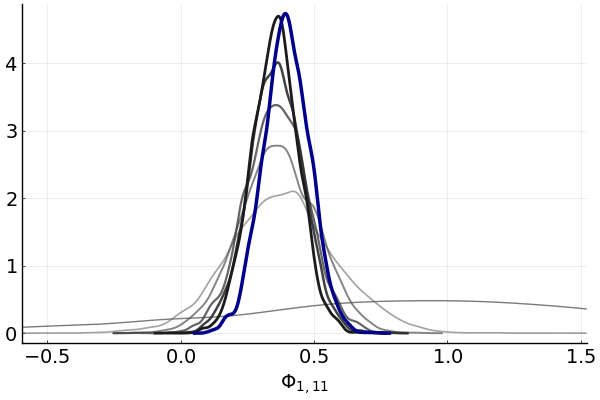}	&
			\includegraphics[width=0.31\textwidth]{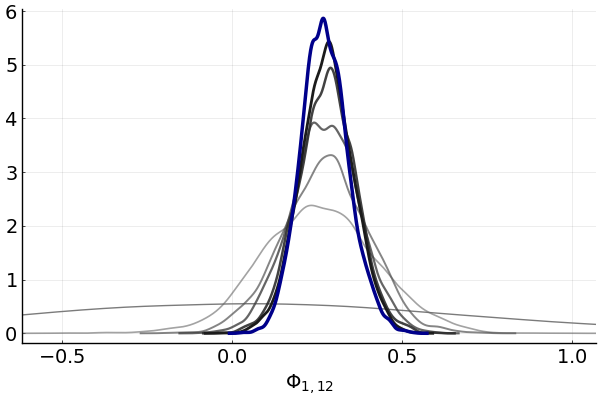} &
			\includegraphics[width=0.31\textwidth]{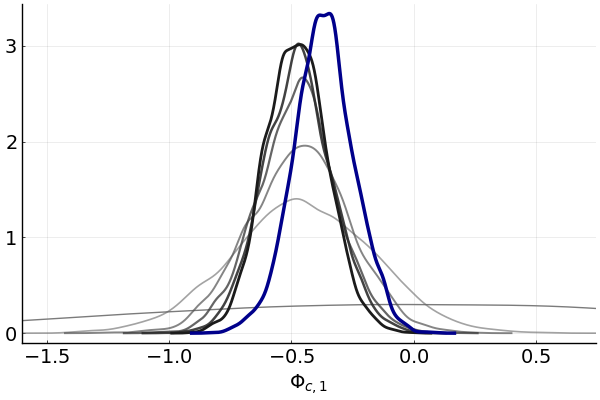} \\
			\includegraphics[width=0.31\textwidth]{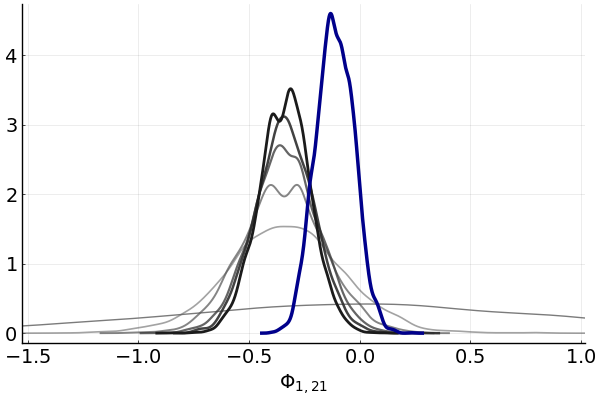}	&
			\includegraphics[width=0.31\textwidth]{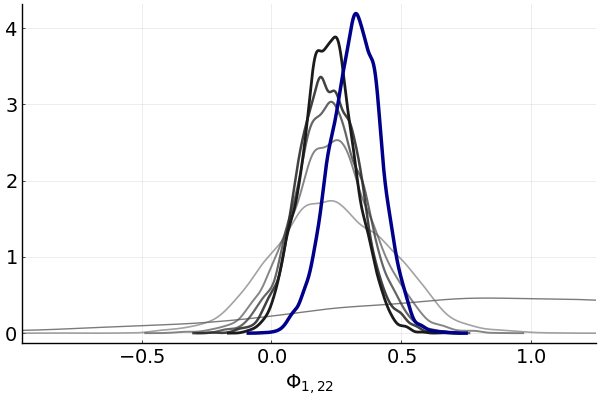} &
			\includegraphics[width=0.31\textwidth]{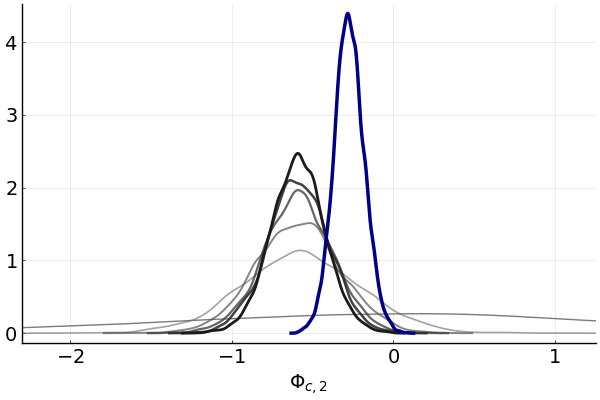} \\
			\includegraphics[width=0.31\textwidth]{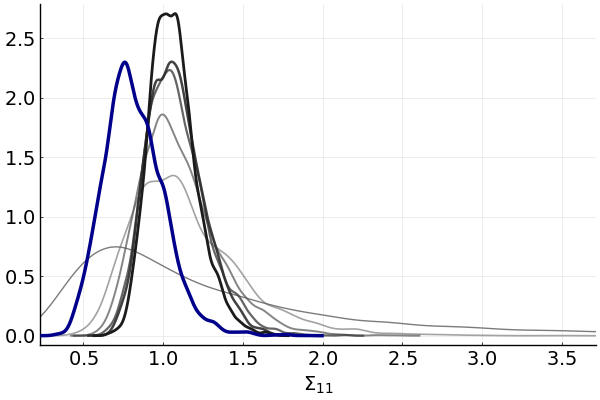}	&
			\includegraphics[width=0.31\textwidth]{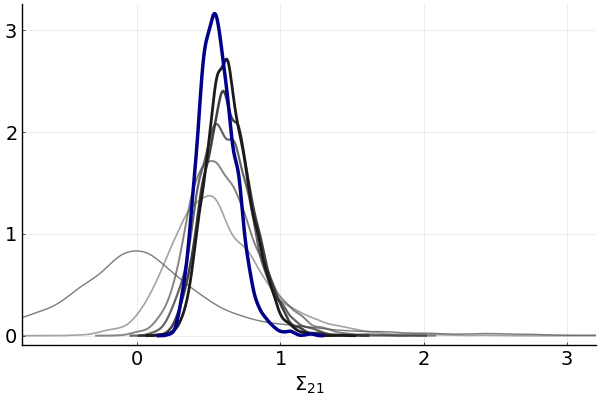} &
			\includegraphics[width=0.31\textwidth]{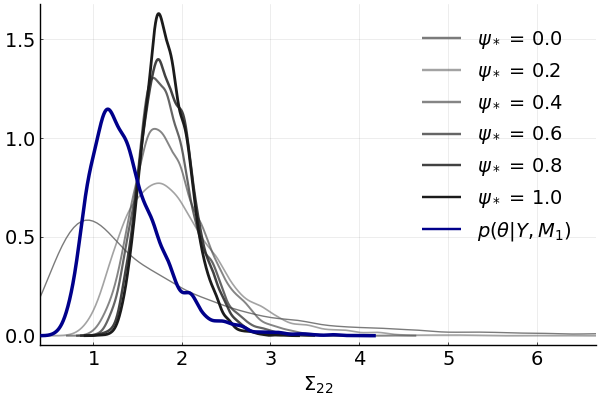} \\
		\end{tabular}
	\end{center}
	{\footnotesize {\em Notes:} Each plot refers to a different parameter. The approximating posterior densities obtained from the tempered $M_0$ likelihood function for $\psi_* \in \{0.0,0.2,0.4,0.6,0.8,1.0\}$ are plotted in shades (the larger $\psi_*$ the darker) of gray. The $M_1$ posterior is depicted in blue. The stochastic volatility parameters $\rho_i, \xi_i, \; i = 1,2$ are not displayed because model $M_0$ is uninformative for them.}\setlength{\baselineskip}{4mm}
\end{figure}

\clearpage

\begin{figure}[h!]
	\caption{VAR-SV: Monte Carlo Approximations of Posterior Statistics for DGP 1}
	\label{appfig:sameres_post}
	\begin{center}
		\begin{tabular}{cc}
			\includegraphics[width=0.45\textwidth]{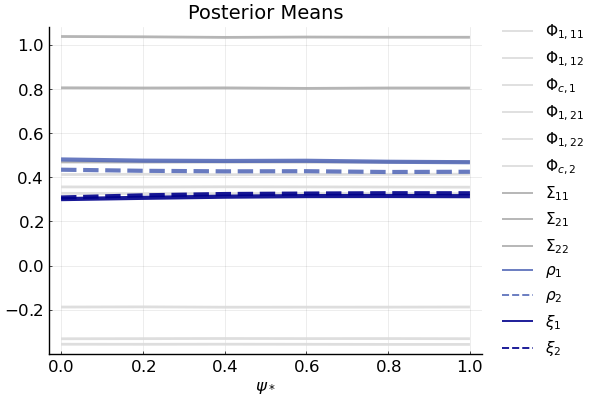}	&
			\includegraphics[width=0.45\textwidth]{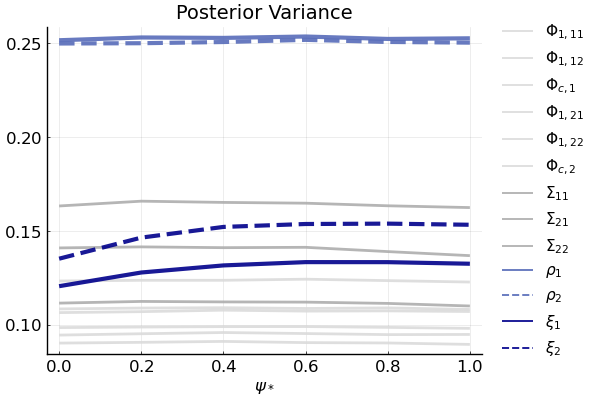} \\
			\includegraphics[width=0.45\textwidth]{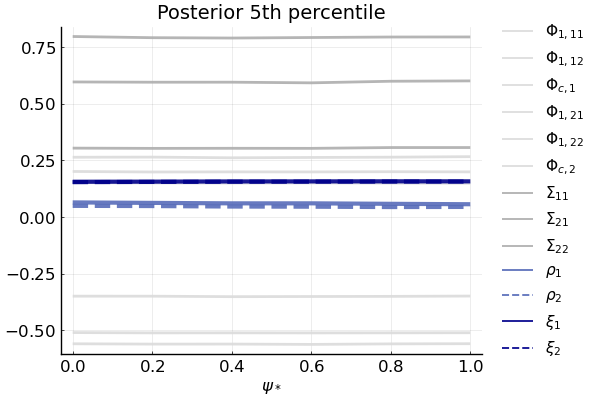} &
			\includegraphics[width=0.45\textwidth]{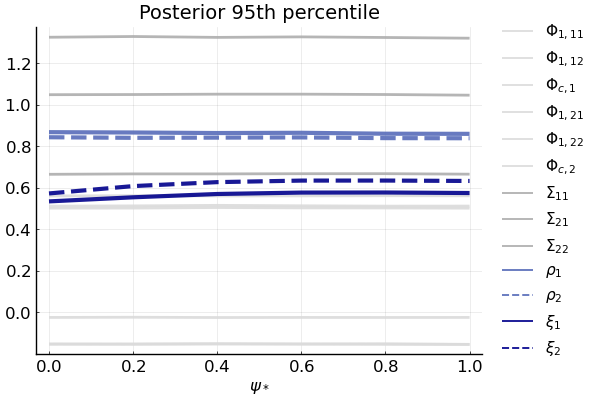}
		\end{tabular}
	\end{center}
	{\footnotesize {\em Notes:} Each panel shows the Monte Carlo approximation of the respective posterior statistic as a function of the tempering parameter $\psi_*$ for the approximating model. Depicted are means across $N_{run}=200$ runs.}\setlength{\baselineskip}{4mm}
\end{figure}

\clearpage

\begin{figure}[t!]
	\caption{VAR-SV: Runtime and Tempering Schedule}
	\label{appfig:runtime.schedule.all}
	\begin{center}
		\begin{tabular}{ccc}
			& Runtime & Tempering Schedules \\
			\rotatebox{90}{\hspace*{0.8in} DGP 1} &
			\includegraphics[width=0.45\textwidth]{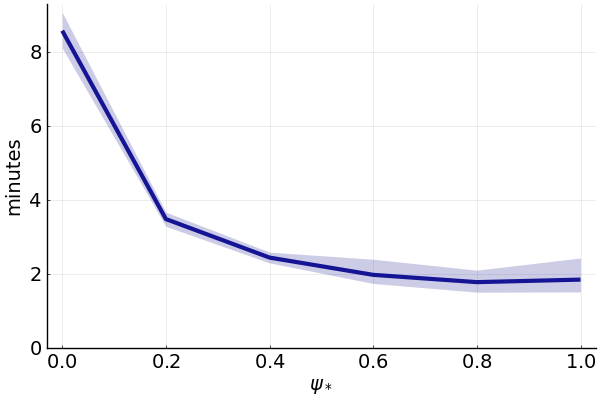}	&
			\includegraphics[width=0.45\textwidth]{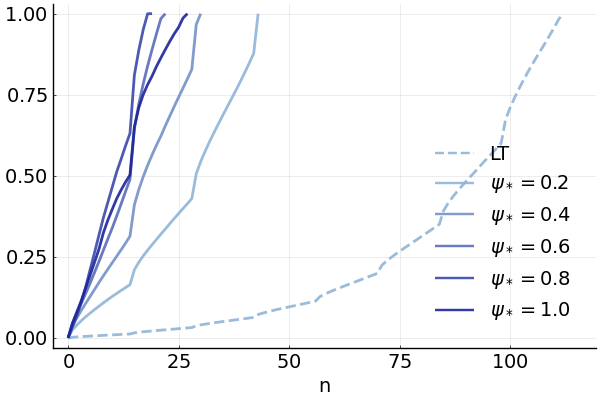} \\
			\rotatebox{90}{\hspace*{0.8in} DGP 2} &
			\includegraphics[width=0.45\textwidth]{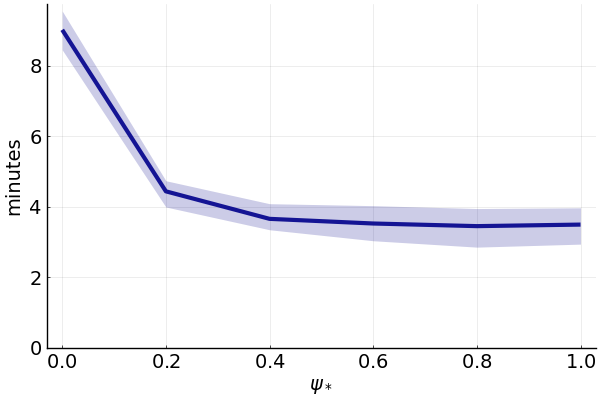}	&
			\includegraphics[width=0.45\textwidth]{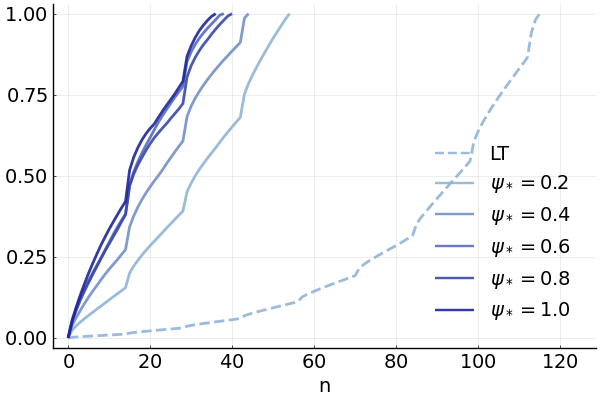} \\
     		\rotatebox{90}{\hspace*{0.8in} DGP 3} &
			\includegraphics[width=0.45\textwidth]{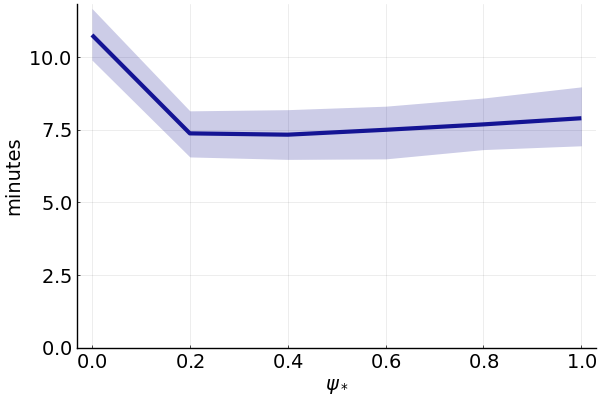}	&
            \includegraphics[width=0.45\textwidth]{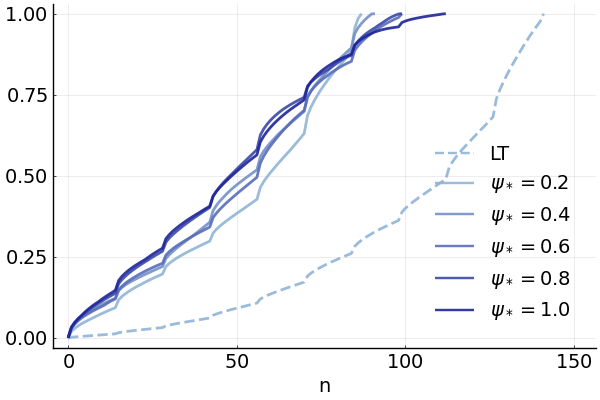}
		\end{tabular}
	\end{center}
	{\footnotesize {\em Notes:} The left panel shows the mean runtime and 90\% confidence interval across $N_{run}=200$ runs. The right panel illustrates the evolution of the tempering schedule by plotting the median value of the tempering parameter at each stage $n$.}\setlength{\baselineskip}{4mm}
\end{figure}

\clearpage

\section{Illustration 3: A Nonlinear DSGE Model}
\label{appsec:DSGE}

\subsection{Equilibrium Conditions, Steady State, and Log-linearization}
\label{appsubsec:DSGE.FOC.Linearization}

We write the social planner's problem stated in the main text as
\begin{eqnarray}
V(K,S) &=& \underset{C,L,K'}{\max} \; u(B,C,L) + \beta \mathbb{E}_{S'|S} [V(K',S')]  \nonumber \\
&  \text{s.t.} &  C + I + K \Phi(K'/K) = Y, \label{appeq:MC} \\
& & Y = f(Z,K,L), \label{appeq:Y} \\
& & I = K' - (1-\delta)K \; . \label{appeq:I}
\end{eqnarray}
We use the following functional forms:
\begin{eqnarray*}
u(B,C,L) &=&  \frac{C^{1-\tau}-1}{1-\tau} - B \frac{L^{1+1/\nu}}{1+1/\nu} , \\
f(Z,K,L) &=& Z K^\alpha L^{1-\alpha}, \\
\Phi(K'/K) &=&  \phi_1 \left( \frac{\exp(-\phi_2(K'/K-1))+ \phi_2(K'/K-1)-1}{\phi_2^2} \right).
\end{eqnarray*}
The exogenous processes evolve according to:
\begin{eqnarray}
Z &= Z_*e^{\hat z}, \quad \hat{z}' = \rho_z \hat{z} + \sigma_z \varepsilon_z' , \nonumber \\
B &= B_*e^{\hat b}, \quad \hat{b}' = \rho_b \hat{b} + \sigma_b \varepsilon_b' \; . \nonumber
\end{eqnarray}
Throughout this section we use $f_i(\cdot)$ to denote the derivative of a function $f(\cdot)$ with respect to its $i$'th argument.

\subsubsection{First-Order Conditions (FOCs)}

Substitute (\ref{appeq:Y}) and (\ref{appeq:I}) into (\ref{appeq:MC}) and then
take FOCs with respect to $L$ and $K'$. The FOC for $L$ takes the form
\[
   u_2(B,C,L)f_3(Z,K,L) + u_3(B,C,L) = 0.
\]
Using the functional forms, this leads to
\be
  (1-\alpha)\frac{Y}{L} = B C^{\tau} L^{1/\nu}.
  \label{appeq:FOC.L}
\ee
Now write
\[
   C = ZK^\alpha L^{1-\alpha} - K' +(1-\delta) K - K \Phi(K'/K).
\]
The FOC for $K'$ takes the form:
\[
  -u_2(B,C,L) \big[ 1+\Phi_1(K'/K) \big] + \beta \mathbb{E} \left[ V_1(K',S') \right] = 0.
\]
Plugging in the expressions for $u_2(\cdot)$ and $V_1(\cdot)$ we obtain
\begin{eqnarray}
\lefteqn{C^{-\tau} \big[ 1+\Phi_1(K'/K) \big]} \label{appeq:FOC.K} \\
& =& \beta \mathbb{E} \left[ C'^{-\tau} \left( \alpha \frac{Y'}{K'} + 1-\delta -  \Phi(K''/K') + \Phi_1(K''/K')\frac{K''}{K'} \right) \right], \nonumber
\end{eqnarray}
where
\be
   \Phi_1(x) = \frac{\phi_1}{\phi_2} \left[ 1 - \exp\{ -\phi_2(x-1)\} \right].
   \label{appeq:Phi1}
\ee

\subsubsection{Steady State}

Rather than taking $(Z_*,B_*)$ as given and solving for $(Y_*,L_*)$ and the remaining steady states, we proceed in the other direction and solve for $(Z_*,B_*)$ as a function of $(Y_*,L_*)$. Notice that the adjustment costs are zero in steady state because $\Phi(1)=0$. Moreover, $\Phi_1(1)=0$. We deduce from (\ref{appeq:FOC.K}) that
\[
   \frac{1}{\beta} = \alpha \frac{Y_*}{K_*} + (1-\delta),
\]
which implies that
\be
   K_* = \frac{\alpha}{1/\beta - (1-\delta)} Y_*.
   \label{appeq:SS.K}
\ee
The capital accumulation equation implies that
\be
    I_* = \delta K_* = \frac{\alpha \delta}{1/\beta - (1-\delta)} Y_*.
    \label{appeq:SS.I}
\ee
The aggregate resource constraint implies that
\be
   C_* = Y_* - I_* = \left( 1- \frac{\alpha \delta}{1/\beta - (1-\delta)} \right) Y_*.
   \label{appeq:SS.C}
\ee
The production function can be solved for $Z_*$:
\be
   Z_* = \frac{Y_*}{K_*^\alpha L_*^{1-\alpha}} = \left(\frac{1/\beta - (1-\delta)}{\alpha}\right)^\alpha \left(\frac{Y_*}{L_*} \right)^{1-\alpha}.
   \label{appeq:SS.Z}
\ee
Finally, we solve (\ref{appeq:FOC.L}) for $B$ to obtain $B_*$:
\[
   B_* = (1-\alpha)\frac{Y_*}{L_*} C_*^{-\tau} L_*^{-1/\nu}.
\]
In the numerical illustration we set $Y_*=L_*=1$.

\subsubsection{Log-Linearization}

Log-linearizing Equations~(\ref{appeq:MC}), (\ref{appeq:Y}), (\ref{appeq:I}), and (\ref{appeq:FOC.L}) yields:
\begin{eqnarray}
\hat y &=& \frac{C_*}{Y_{*}} \hat c + \frac{I_*}{Y_{*}} \hat i \label{appeq:lin.MC} \\
\hat y &=& \hat{z} + \alpha \hat k + (1-\alpha) \hat l \label{appeq:lin.Y} \\
\delta \hat i &=& \hat k' - (1-\delta) \hat k \label{appeq:lin.I} \\
(1+1/\nu)\hat l &=&  \hat{y} - \hat{b} - \tau \hat c. \label{appeq:lin.L}
\end{eqnarray}

We proceed with the log-linearization of $\Phi_1(x)$ in (\ref{appeq:Phi1}). Differentiating with respect to the argument yields
\[
   \Phi_{11}(x) = \phi_1 \exp \{ - \phi_2(x-1)\}.
\]
Log-linearizing around $x=\exp(z)=1$ leads to the approximation:
\[
  \Phi_1\big(\exp(z)\big) \approx \Phi_1(1) + \Phi_{11}(1)\cdot 1 \cdot (z-0).
\]
In turn, we can write
\[
   \Phi_1(K'/K) \approx \phi_1(\hat{k}'-\hat{k}),
\]
which shows that the linex adjustment cost function is equivalent, up to second order, to a quadratic adjustment cost function
\[
   \Phi(K'/K)  \approx \frac{\phi_1}{2} \big( K'/K - 1 \big)^2.
\]

We now turn to the log-linearization of (\ref{appeq:FOC.K}) using the observation that $\Phi_1(1)=0$:
\begin{eqnarray*}
 \lefteqn{ -\tau C_*^{-\tau} \hat{c} + C_*^{-\tau} \phi_1(\hat{k}'-\hat{k}) } \\
 &=& - \tau \beta C_*^{-\tau} ( \alpha Y_*/K_* + 1-\delta) \mathbb{E}[\hat{c}'] + \alpha \beta C_*^{-\tau} \frac{Y_*}{K_*} \mathbb{E}[\hat{y}'-\hat{k}']
 + \phi_1 \beta C_*^{-\tau} \mathbb{E}[ \hat{k}'' - \hat{k}']. \nonumber
\end{eqnarray*}
Multiplying by $C_*^{\tau}$, using (\ref{appeq:SS.K}), and noting that $\hat{k}'$ is in the information for the conditional expectation $\mathbb{E}[\cdot]$ yields the simplified equation:
\be
	 -\tau \hat{c} +  \phi_1(\hat{k}'-\hat{k})
	= - \tau \mathbb{E}[\hat{c}'] + \big(1-\beta(1-\delta)\big) \big(\mathbb{E}[\hat{y}']-\hat{k}'\big)
	+ \phi_1 \beta \big(\mathbb{E}[ \hat{k}''] - \hat{k}'\big). \label{appeq:lin.K}
\ee
Equations~(\ref{appeq:lin.MC}) to~(\ref{appeq:lin.K}) and the laws of motion for $\hat{z}$ and $\hat{b}$ form a linear rational expectations system that determines the dynamics of the model.

After setting $Y_*=L_*=1$, the measurement equations in (\ref{eq:DSGE.ME}) can be written  as
\be
  \ln Y^o = \hat y + \eta_Y, \quad
  \ln I^o = \ln \left( \frac{\alpha \delta}{1/\beta - (1-\delta)} \right) + \hat{i} + \eta_I, \quad \ln L^o = \hat{l} + \eta_l.
\ee

\subsection{Model Solution, and Computational Details}
\label{appsubsec:RBC.solution}

While the approximate model $M_0$ refers to a first-order linearization around the steady state and is described in Section \ref{appsubsec:DSGE.FOC.Linearization} above, we obtain $M_1$ as a second-order linearization around the steady state, computed following \cite{SGU2004}. To implement it in Julia, we use the package {\em SolveDSGE}, developed by Richard Dennis and available at {\em https://github.com/RJDennis}.

\subsection{Further Results for the RBC Model}
\label{appsubsec:RBC.moreresults}

\begin{figure}[h!]
	\caption{RBC Model: Simulated Data}
	\label{appfig:RBC.data}
	\begin{center}
      \includegraphics[width=0.98\textwidth]{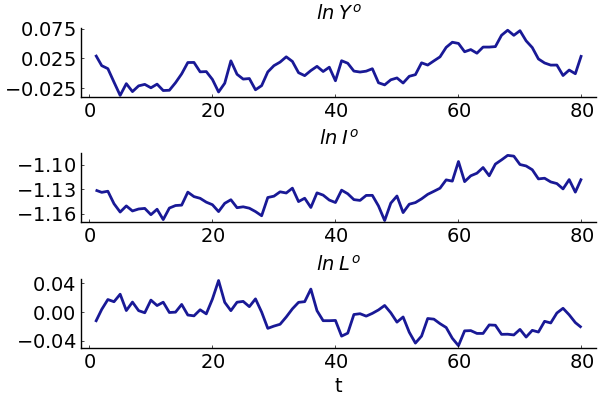}
    \end{center}
\end{figure}

\begin{figure}[h!]
	\caption{RBC Model: Target and Approximate Posterior Densities}
	\label{appfig:RBC.distEvol2D_1}
	\begin{center}
		\begin{tabular}{ccc}
			\includegraphics[width=0.31\textwidth]{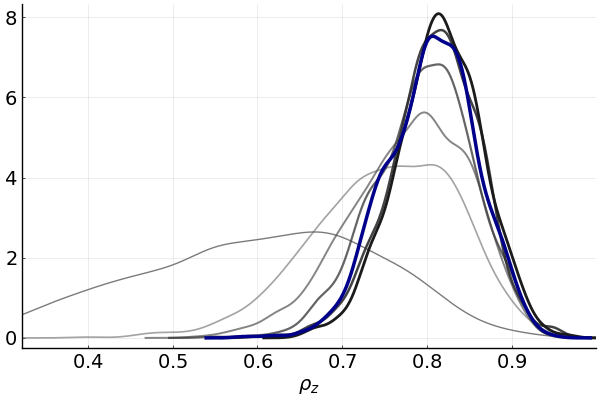}	&
			\includegraphics[width=0.31\textwidth]{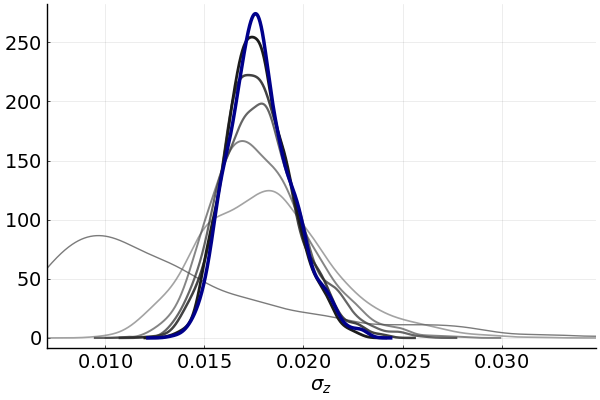} &
			\includegraphics[width=0.31\textwidth]{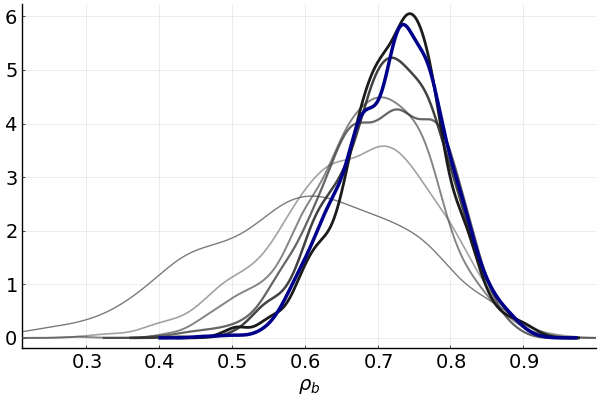} \\
			\includegraphics[width=0.31\textwidth]{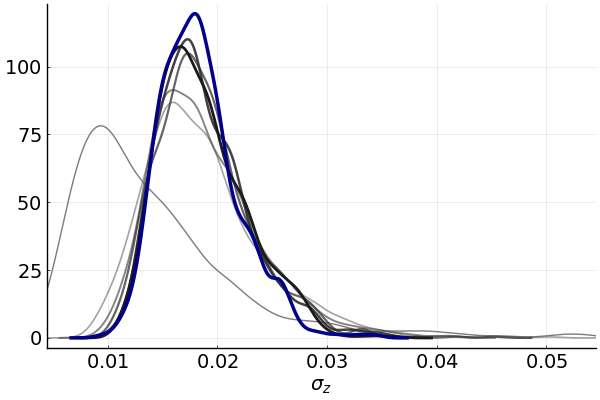}	&
			\includegraphics[width=0.31\textwidth]{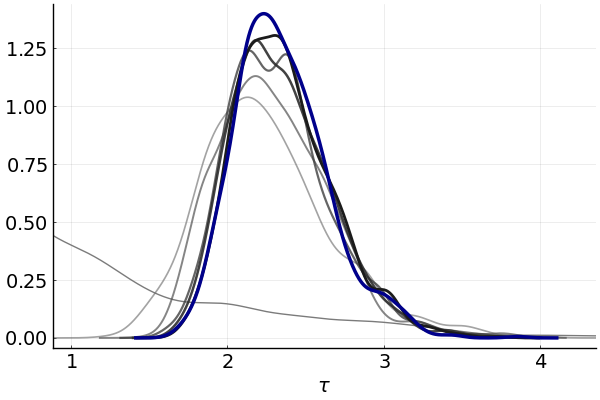} &
			\includegraphics[width=0.31\textwidth]{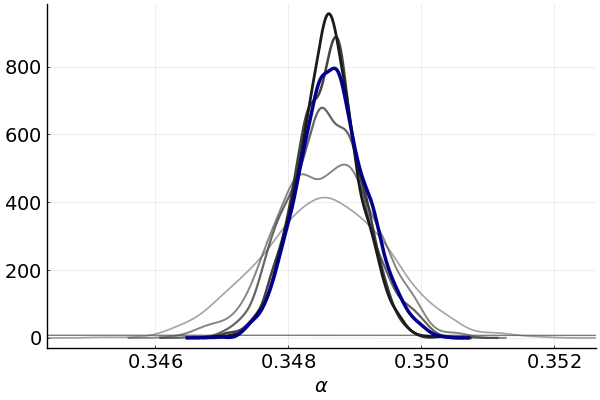} \\
			\includegraphics[width=0.31\textwidth]{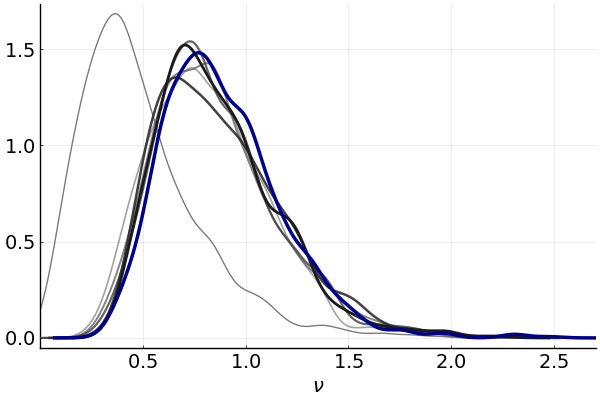}	&
			\includegraphics[width=0.31\textwidth]{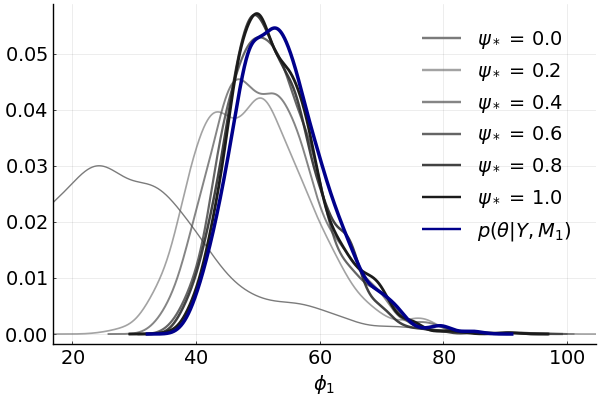}
		\end{tabular}
	\end{center}
	{\footnotesize {\em Notes:} Each plot refers to a different parameter. The approximating posterior densities obtained from the tempered $M_0$ likelihood function for $\psi_* \in \{0.0,0.2,0.4,0.6,0.8,1.0\}$ are plotted in shades (the larger $\psi_*$ the darker) of gray. The $M_1$ posterior is depicted in blue.}\setlength{\baselineskip}{4mm}
\end{figure}

\begin{figure}[t!]
	\caption{RBC Model: Absolute Runtimes}
	\label{appfig:RBC.absruntime}
	\begin{center}
			\includegraphics[width=0.5\textwidth, clip]{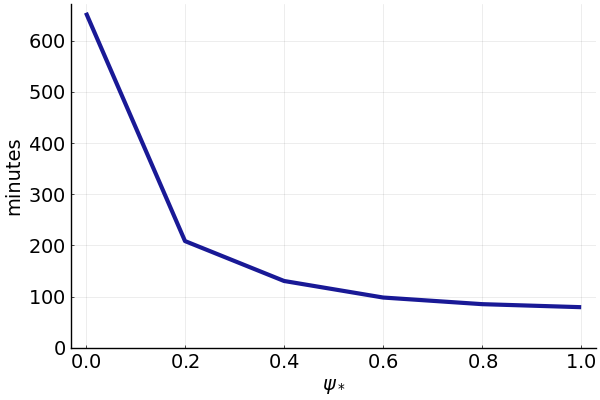}
	\end{center}
	{\footnotesize {\em Notes:} Single run ($N_{run}=1$) }\setlength{\baselineskip}{4mm}
\end{figure}

\end{appendix}

\end{document}